\documentclass[pra,10pt,twocolumn,superscriptaddress]{revtex4-2}

\usepackage{graphicx}
\usepackage[usenames,dvipsnames]{color}
\usepackage{dcolumn}
\usepackage{bm}
\usepackage{hyperref}
\usepackage{amsmath}
\usepackage{braket}
\usepackage{physics}
\usepackage[section]{placeins}
\usepackage{dsfont}

\renewcommand{\ip}[2]{\left\langle #1 | #2\right\rangle}

\begin{document}

\preprint{APS/123-QED}

\title{Effect of repeated projective measurements on a two-qubit system undergoing dephasing\\}

\author{Hammas Hussain Ali}
 \altaffiliation{These authors contributed equally: Hammas Hussain Ali, Muhammad Abdullah Ijaz, Fariha Hassan, and Diya Batool.}
\author{Muhammad Abdullah Ijaz} 
\altaffiliation{These authors contributed equally: Hammas Hussain Ali, Muhammad Abdullah Ijaz, Fariha Hassan, and Diya Batool.}
\author{Fariha Hassan}
\altaffiliation{These authors contributed equally: Hammas Hussain Ali, Muhammad Abdullah Ijaz, Fariha Hassan, and Diya Batool.}
\author{Diya Batool}
\altaffiliation{These authors contributed equally: Hammas Hussain Ali, Muhammad Abdullah Ijaz, Fariha Hassan, and Diya Batool.}

\author{Adam Zaman Chaudhry}
\altaffiliation{adam.zaman@lums.edu.pk}
\affiliation{%
 School of Science and Engineering, Lahore University of Management Sciences (LUMS),
Opposite Sector U, D.H.A, Lahore 54792, Pakistan.
}%



\date{\today}

\begin{abstract}
The entanglement dynamics of an exactly solvable, pure dephasing model are studied. Repeated projective measurements are performed on the two-qubit system. Due to the system-environment interaction, system-environment correlations are established between each measurement. Consequently, the environment state keeps evolving. We investigate the effect of this changing environment state on the entanglement dynamics. In particular, we compare the dynamics with the case where the environment state is repeatedly reset. 


\end{abstract}
%
\maketitle


\section{\label{sec:level1}Introduction }

Rapid repeated measurements can slow down the temporal evolution of quantum systems \cite{10.1063/1.523304}. This is called the Quantum Zeno effect (QZE) and it is of great theoretical and experimental interest, especially in practical implementations of emerging quantum technologies \cite{kwiat1999high, franson2004quantum,Gordon_2010,zhu2014suppressing, chaudhry2014zeno,facchi2010quantum,smerzi2012zeno, Facchi_2000,PhysRevA.83.022107,Zhang_2011,PhysRevLett.110.240403}. Interestingly, if the repeated measurements are not rapid enough, the temporal evolution of the quantum system can accelerate. This is called the Quantum anti-Zeno effect (QAZE) \cite{article,PhysRevLett.86.2699,Koshino_2005}. Both QZE and QAZE have been explored in many setups, such as superconducting qubits, Josephson junctions, and nanomechanical oscillators \cite{Zhang_2011,Slichter_2016,cao2012transition, PhysRevB.81.115307,PhysRevLett.92.200403}. There have been studies of QZE and QAZE for a single qubit system interacting with an environment that evolves as measurements on the system are made. The environment evolves because, especially with strong system-environment coupling strength, the system and the environment can establish significant correlations; as a result, any measurement performed thereafter on the system can significantly affect the environment state \cite{chaudhry2014zeno,khalid2019quantum,khan2021quantum,khan2022generalized}. In this paper, we aim to study the dynamics of a two-qubit system that interacts with its environment. In the presence of repeated projective measurements performed on the system, the environment is repeatedly projected onto a non-equilibrium state. The subsequent system dynamics are then highly non-trivial. We study the effect of this evolving environment state for different numbers of measurements, system-environment coupling strength, and different environment characteristics. 

Entanglement is the fundamental quantity we use to study the dynamics of the central two-qubit system. As is well-known, entanglement is a crucial quantum-mechanical feature that enables the use and study of non-classical physics \cite{nielsen2010quantum,beck2012quantum,waseem2020quantum}. The measure for entanglement used within this study is concurrence \cite{wootters1998entanglement, qi2017measuring, Schlosshauer:2014pgr}. We model the environment of the two-qubit system as a collection of simple harmonic oscillators. Considering the dephasing timescales to be much shorter than dissipation, we consider only a pure dephasing type interaction between the system and the environment \cite{chaudhry2014zeno, chaudhry2016general, chaudhry2017quantum, mirza2023role,Schlosshauer:2014pgr,pet}. Such a study of the effect of the changing environment state on the two-qubit state is especially important as two-qubit systems can display physics (such as the entanglement between the qubits) entirely absent from single-qubit systems.


This paper is organized as follows. In Sec.~\ref{sec:level2}, we introduce the two-qubit dephasing model we use and discuss its solution for the case where the environment state is repeatedly reset after each measurement. We then examine analytically the effect of the changing environment state. In Sec.~\ref{sec:level3}, we present results for the entanglement dynamics for the case where the environment changes upon measurements and compare this evolution to the case where the environment is repeatedly reset. We also look at the effect of using different measurement intervals on the evolution of the entanglement. Finally, we present our conclusions in Sec.~\ref{sec:conc}. Technical details of the calculations are shown in the appendices.


\section{\label{sec:level2}Mathematical Framework:\protect\\ }

\subsection{\label{sec:level21}Pure dephasing Model}
We examine a pure dephasing model with the system comprised of two qubits. This means that the system and environment do not exchange energy, and there is decoherence without dissipation. The two qubits interact with a common environment of a collection of harmonic oscillators \cite{chaudhry2014zeno}. The total Hamiltonian, $H$, is comprised of the Hamiltonian of the system, environment, and the interaction (we use dimensionless units throughout with $\hbar$ = 1),
 \begin{equation} \label{eq:1}
     H= H_S+H_E+H_{I},
 \end{equation} 
where $H_S=\frac{\omega_0}{2}(\sigma_z^{ (1)}+\sigma_z^{(2)})$, $H_E= \sum_r \omega_r b_r^\dagger b_r$, and $H_I= (\sigma_z^{ (1)}+\sigma_z^{(2)})\sum_r(g_r^* b_r+g_r b_r^\dagger) $. Here $g_r$ is the coupling strength between the $r^{\text{th}}$ mode of the environment with frequency $\omega_r$ and the two-qubit system, and $\omega_{0}$ is the energy difference between the ground and excited state of the two-level systems. We use the eigenstates of $\sigma_z^{(1)}$ and $\sigma_z^{(2)}$ $\ket{k,l}$, defined as 
\begin{align*} 
    \sigma_z^{(1)}\ket{k,l}=k \ket{k,l}, \\
    \sigma_z^{(2)}\ket{k,l}=l \ket{k,l},
\end{align*} 
with $k,l = \pm 1$. We consider the case in which the system and the environment initially form a product state \cite{mirza2023role},
\begin{align*} 
    \rho(0) &=\rho_S (0) \otimes \rho_E(0), \\
            &= \ket{\psi}\bra{\psi} \otimes \frac{e^{-\beta H_E}}{Z_E},
\end{align*}
where $ Z_E =  \text{Tr}_E [e^{-\beta H_E}] $ and we assume that the system is initially in the pure state $\ket{\psi}$ and the environment is in a thermal equilibrium state. Because we have a pure dephasing model, the off-diagonal elements of the system density matrix are of primary interest. The density matrix takes the form (see Appendices \ref{appendix} and \ref{appendb}) 
\begin{align} \label{eq:3}
[\rho_S(t)]_{k'l',kl} &= [\rho_S(0)]_{k'l',kl}  e^{-i\frac{\omega_0}{2}(k'+l'-k-l)t} \nonumber \\ &\times e^{-i\frac{\Delta(t)}{2}(k'l'-kl)} e^{-\frac{1}{4}(k+l-k'-l')^2\gamma(t)}.
\end{align}
Here $[\rho_S(0)]_{k'l', kl}$ are the matrix elements of the initial system state and 
\begin{align*}
    \gamma(t)= \sum_r \frac{4|g_r|^2}{\omega_r^2}[1-\cos(\omega_r t)]\coth(\frac{\beta \omega_r}{2})
\end{align*} 
is the environment-induced dephasing factor. On the other hand,
\begin{align*}
    \Delta (t)=\sum _r \frac{4|g_r|^2}{\omega_r^2}[\sin(\omega_r t)-\omega_r t],
\end{align*}
quantifies the indirect interaction between the qubits via the common environment. To calculate the effect of the environment, we replace the summation $\sum_{r} {|g_{r}|}^{2}(\hdots)$ with the integral  $\int \, d\omega \, J(\omega)  (\hdots)$ \cite{pet}. Here, we take the spectral density to be $J(\omega)= G \omega^s \omega_c^{1-s}$, where $\omega_{c}$ is the cutoff frequency of the environment, and $s$ is the Ohmicity parameter. $s<1$ represents the sub-Ohmic regime, $s>1$ is the super-Ohmic regime, and $s=1$ is the Ohmic regime. Our primary quantity of interest in this paper is the entanglement between the two qubits comprising the system. To quantify this, we use the concurrence defined as \cite{wootters1998entanglement}, 
\begin{align} \label{eq:2}
    R = \sqrt{\sqrt{\rho_S} \tilde{\rho} _S\sqrt{\rho_S}},
\end{align}
where
\begin{align*} 
    \tilde{\rho}_S = (\sigma_{y} \otimes \sigma_{y}) \rho_S (\sigma_{y} \otimes \sigma_{y}),
\end{align*}
and the concurrence is given by $C(\rho_S) = \text{max}\{0, \lambda_1 -\lambda_2 -\lambda_3 -\lambda_4 \}$. $\lambda_i$ are the eigenvalues of the Hermitian matrix $R$ in descending order.

\subsection{\label{sec:level22} Repeated Projective Measurements}

In this section, we aim to incorporate repeated projective measurements on the system. It is important to note that, after each measurement, the environment is generally not in a thermal equilibrium state. We consider the time interval between the equally spaced measurement to be $\tau$. We find that the density matrix of the system and the environment after the $(N-1)^{\text{th}}$ and before the $N^{\text{th}}$  measurement is given by 
\begin{widetext}
\begin{align} \label{eq:4}
     \rho \big [(N-1)\tau \leq t < N\tau  \big ]&= \frac{1}{Z_{N-1}} U [t-\tau(N-1)] \big[P_\psi U (\tau) \big]^{N-1}\rho(0) \big [U^\dagger (\tau) P_\psi \big]^{N-1} U^\dagger [t-\tau(N-1)], \\
    \label{eq:5}
    Z_{N-1} &= \text{Tr}_{S,E}\bigg\{ \big[P_\psi U (\tau) \big]^{N-1}\rho(0) \big [U^\dagger (\tau) P_\psi \big]^{N-1} U^\dagger \big[t-\tau(N-1)\big]  \bigg\}.
\end{align}
\end{widetext}
We can define $t'=t-\tau(N-1)$ for brevity and convenience. Here, $U(t)$ is the combined time-evolution operator for the system and the environment, and $P_\psi$ is the projection operator for the measurement. Now, we find that the time evolution operator $U(t)$ can be written as
\begin{align} \label{eq:6}
    U(t)= U_0(t) U_I(t)
    =e^{-i\frac{\omega_0}{2}(\sigma_z^{(1)}+\sigma_z^{(2)})t} e^{- i H_E t} U_I(t),
\end{align}
where $U_I(t)$ is the time-evolution operator associated with the interaction Hamiltonian (for details, see Appendix \ref{appendix}). Using this time-evolution operator in Eq.~\eqref{eq:4}, with the initial system-environment state $\rho(0)=\ket{\psi}\bra{\psi} \otimes \rho_E(0)$,  where $\rho_E(0)$ is the initial thermal equilibrium state of the environment, we can massage the density matrix to the form
\begin{equation} \label{eq:7}
[\rho_S(t)]_{k'l',kl}=\frac{e^{\frac{i}{2}\Delta(t')(kl-k'l')}}{Z_{N-1}} \text{T}_{\text{env}} \text{T}_{\text{sys}}, 
\end{equation}
where
\begin{align*}
    \text{T}_{\text{env}} =& \text{Tr}_{E} \big \{  \big[\bra{\psi} U (\tau) \ket{\psi} \big ]^{N-1}\rho_{E}(0) \\
    & \times \big[\bra{\psi} U^\dagger (\tau) \ket{\psi} \big ]^{N-1}  e^{-R_{kl,k'l'}(t')} \big \}, \\
    \text{T}_{\text{sys}} =& \text{Tr}_{S} \big \{P_\psi P_{kl,k'l'}\}.
\end{align*}
We have defined $P_{kl,k'l'}= \ket{k,l}\bra{k',l'}$. Moreover,
\begin{align*}
    R_{kl,k'l'}(t')=\frac{1}{2}(k+l-k'-l')\sum_r [\alpha_r(t')b_r^\dagger-\alpha^*_r(t')b_r],
\end{align*}
where
\begin{align*}
    \alpha_r(t')= \frac{2g_r(1-e^{i\omega_r t'})}{\omega_r}.
\end{align*}
This result can be simplified as detailed in Appendix \ref{appendc}. We can then study the dynamics of the system in the presence of repeated measurements. It must be noted that the environment state keeps changing throughout. This change is encapsulated in the form of the phase factors $\gamma_{ij}, \mu_{ij}, \epsilon_{ij}$ and $\sigma_{ij}$ which emerge from $T_{\text{Env}}$ (see Appendix \ref{appendc}). The complete form of $[\rho_S(t)]_{{k'l',kl}}$ containing these factors is shown in Eq.~\eqref{eq:FinalExpanded}. These factors accumulate with the number of measurements and show dependence on the time interval between measurements. This is to be contrasted with the scenario where we assume that the environment state is reset after each measurement on the system. In such a case, the system dynamics are given simply by Eq.~\eqref{eq:3}.

\section{\label{sec:level3} Entanglement Dynamics}

Until now, we have presented the density matrix of the two qubits when the environment is repeatedly reset and when it is not. In the current section, we shall see the entanglement dynamics with these two different density matrices.

\subsection{\label{sec:level31} Projecting onto a Product State of Two Qubits}
To begin, we consider the projection operator $P_\psi = \ket{\psi}\bra{\psi}$ with
\begin{align} \label{eq:productstate}
\ket{\psi} &= \ket{1}_X\ket{1}_X \\
&= \frac{1}{2} \big(\ket{-1,-1} + \ket{1,1} + \ket{-1,1} + \ket{1,-1}\big) \nonumber
\end{align}
Here, $\sigma_x \ket{1}_X = \ket{1}_X$. This is a product state. After each measurement, the system will then have zero concurrence. With this projective measurement, we shall compare the concurrence dynamics for the case where the environment is reset and when it is not. 

\begin{figure}[]
\begin{center}
  \includegraphics[scale=0.45]{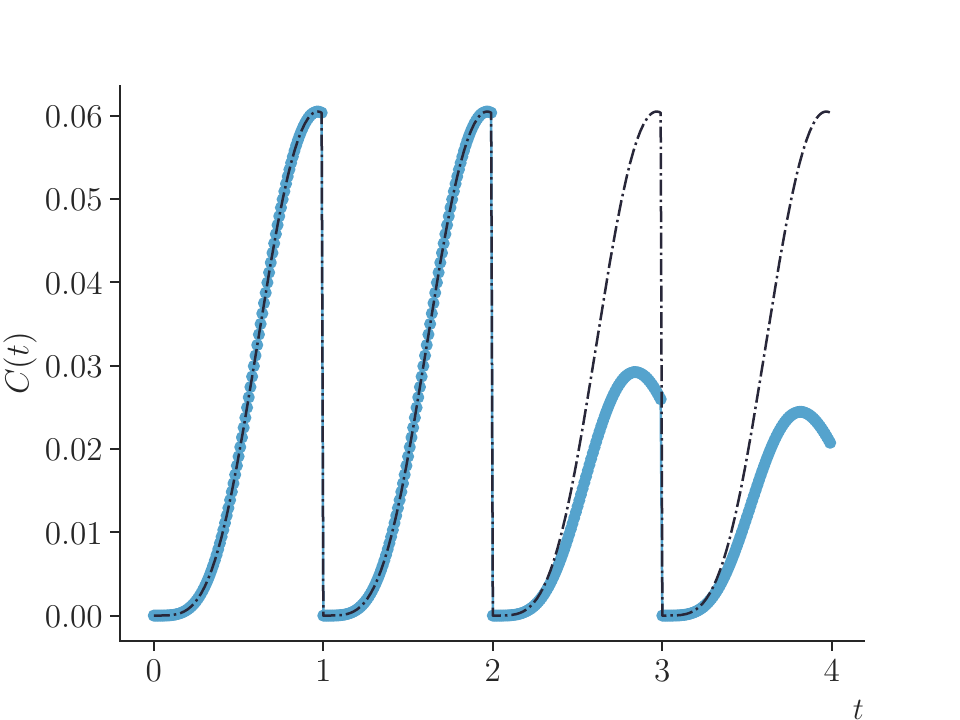}\\
  \caption{The black dot-dashed curve represents the entanglement dynamics when the environment remains unchanged upon a measurement of the system. The blue circles show entanglement dynamics for an environment that changes upon a measurement of the system. Note that the curve repeatedly drops to zero; a measurement is made at these points, and the state is reset to the initial product state. Here, we have a sub-Ohmic environment with $s= 0.1$ and strong system-environment coupling ($G = 1$). Our measurement interval is $\tau = 1$. Throughout, we use dimensionless units with $\hbar = 1$, and the cutoff frequency has been set to $\omega_c = 2$.  }\label{fig1}
  \end{center}
\end{figure}

\begin{figure}[]
\begin{center}
  \includegraphics[scale=0.47]{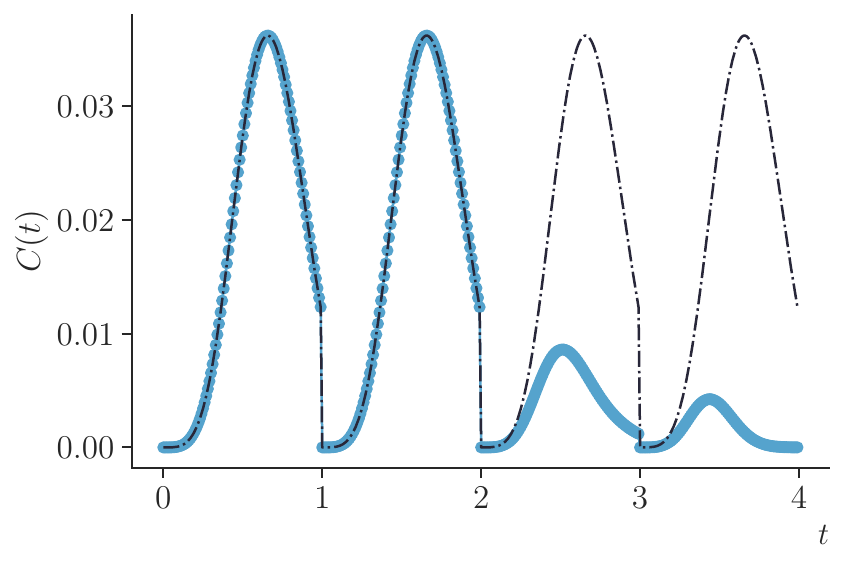}
  \caption{Dynamics with the same parameters as Fig.~\ref{fig1} except that now $G = 2$.}\label{fig_pg}
  \end{center}
\end{figure}

\begin{figure}[]
\begin{center}
  \includegraphics[scale=0.45]{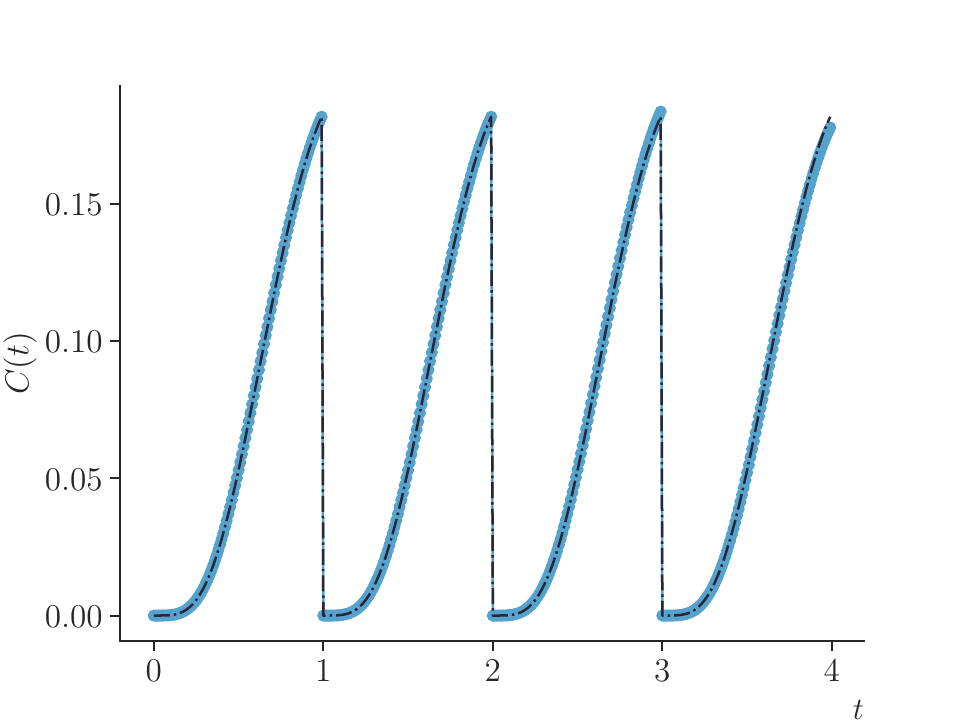}\\
  \caption{We now have an Ohmic environment with $s = 1$. The rest of the parameters are the same as Fig.\ref{fig1}. }\label{fig3}
  \end{center}
\end{figure}

\begin{figure}[]
\begin{center}
  \includegraphics[scale=0.45]{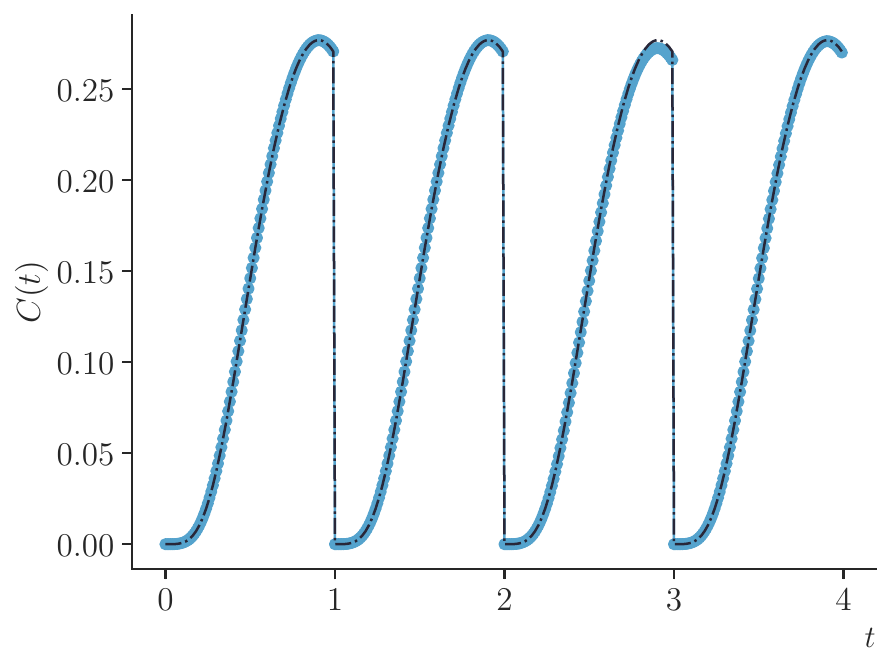}\\
  \caption{Same as Fig.~\ref{fig1}, except that we now have a super-Ohmic environment with $s = 2$.}\label{fig4}
  \end{center}
\end{figure}

It is clear from Fig.~\ref{fig1} that after only two measurements, the changing environment state can already have an evident influence on the system dynamics. In Fig.~\ref{fig_pg}, the coupling strength is doubled. As expected, the impact of the changing environment on system dynamics is enhanced. 
In Fig.~\ref{fig3}, we move from the sub-Ohmic regime to the Ohmic regime. In this case, the changing environment has a much smaller effect. This is because of the shorter correlation time of the environment in the case of an Ohmic environment. As expected, for a super-Ohmic environment, the effect of the changing environment should also be negligible. This is indeed the case, as illustrated in Fig.~\ref{fig4}. 

\subsection{\label{sec:level32}Projecting Onto a Maximally Entangled State of Two Qubits}

For this section, we use the projection operator $P_\psi$ with  
\begin{align}
\ket{\psi} = \frac{1}{\sqrt{2}}{(\ket{-1,-1} + \ket{1,1})}
\end{align}
This maximally entangled state allows us to observe how concurrence decays with time. We can compare the dynamics of concurrence for the case of the environment being reset with the case when it is not. 

\begin{figure}[]
\begin{center}
  \includegraphics[scale=0.45]{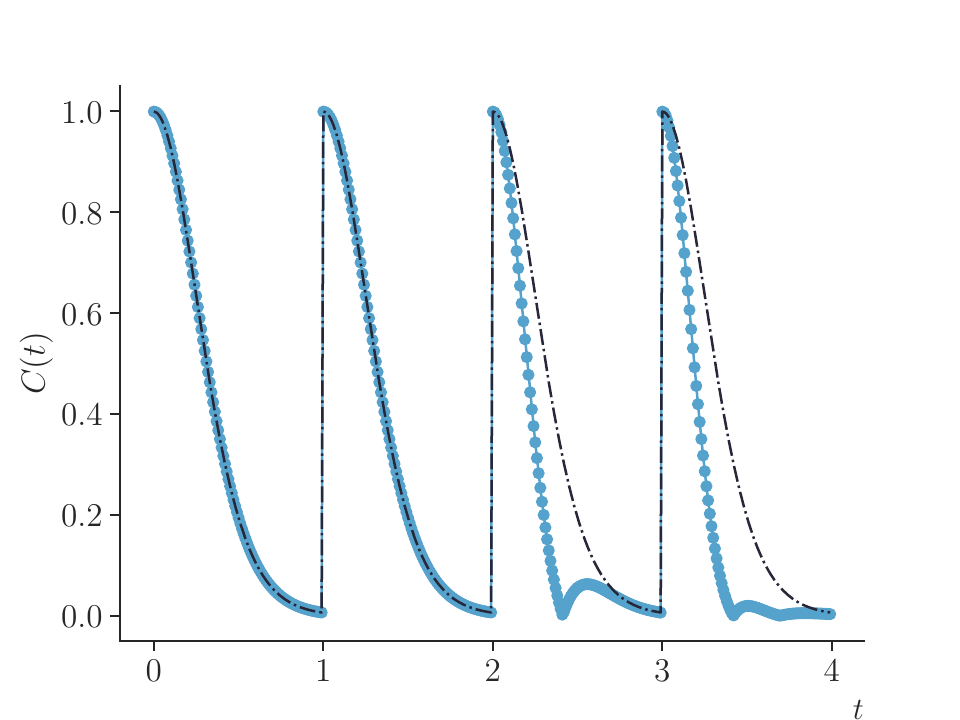}\\
  \caption{As in Fig.~\ref{fig1}, we are in the strong coupling regime with coupling strength $G = 1$. We are also in the sub-Ohmic regime with $s = 0.1$ and our measurement interval is $\tau = 1$.}\label{fig5}
  \end{center}
\end{figure}

\begin{figure}[]
\begin{center}
  \includegraphics[scale=0.49]{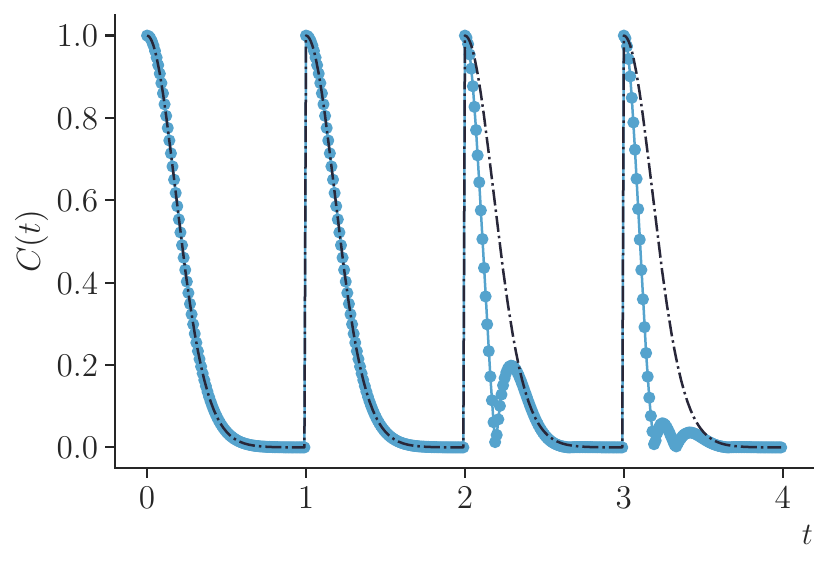}
  \caption{Entanglement dynamics with the same parameters as Fig.~\ref{fig5}  except that now $G = 2$.}\label{fig_bg}
  \end{center}
\end{figure}

\begin{figure}[]
\begin{center}
  \includegraphics[scale=0.45]{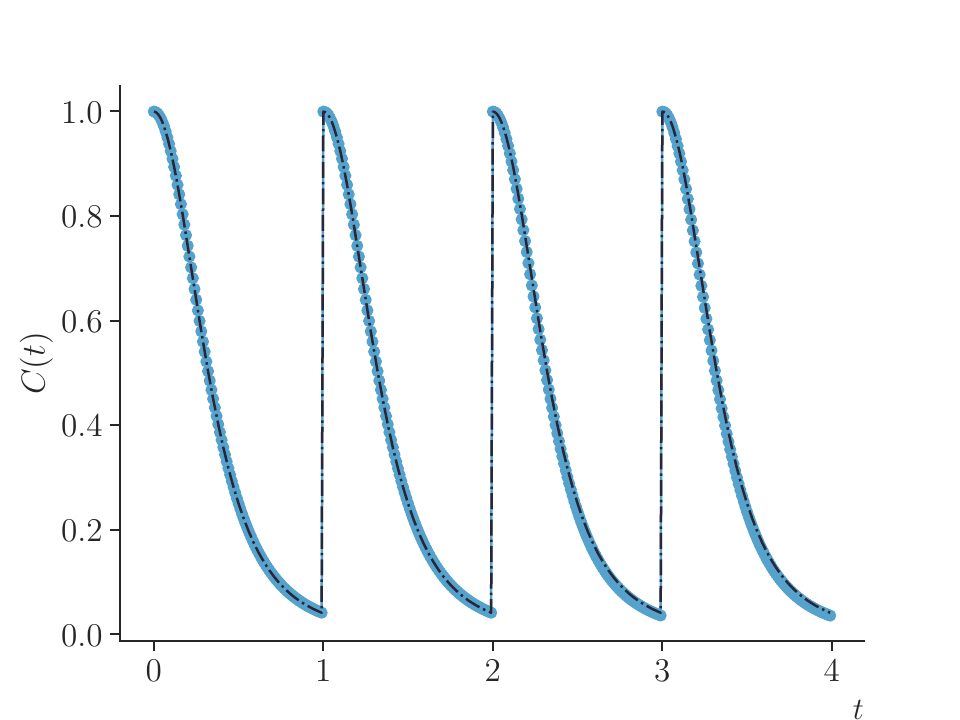}\\
  \caption{This is the same as Fig.~\ref{fig5} except that now $s = 1$.}\label{fig7}
  \end{center}
\end{figure}

\begin{figure}[]
\begin{center}
  \includegraphics[scale=0.45]{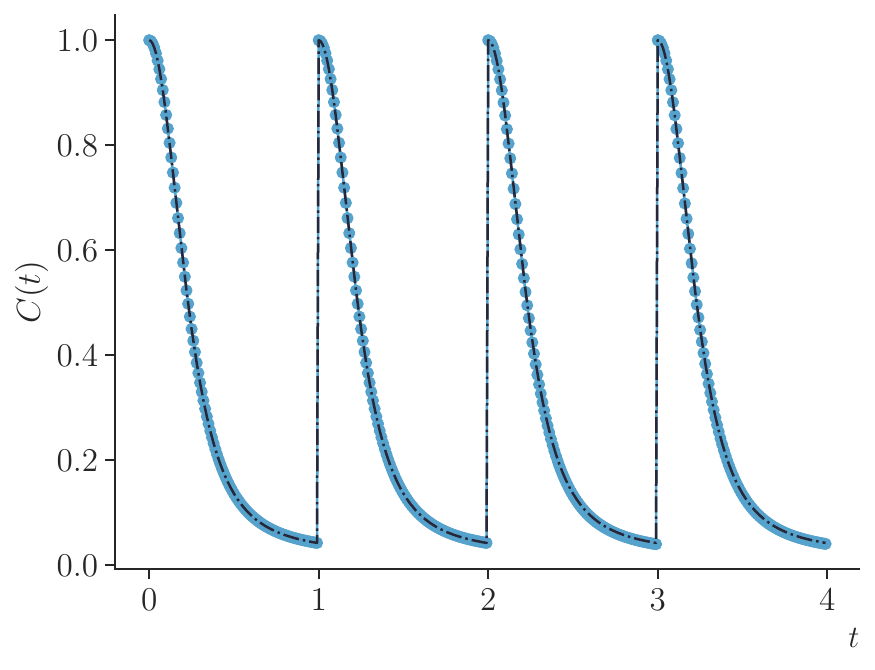}\\
  \caption{This uses the same parameters as Fig.~\ref{fig5}, except that now $s = 2$.}\label{fig8}
  \end{center}
\end{figure}

In Fig.~\ref{fig5}, we see that in the strongly coupled, sub-Ohmic regime, the concurrence decays faster in the case of a changing environment after the second measurement. In fact, with the changing environment state, we can see a much more rapid death of the entanglement followed by a rebirth. In Fig.~\ref{fig_bg}, when the coupling strength is doubled, the influence of a changing environment is enhanced. In Fig.~\ref{fig7}, we move to the Ohmic regime and see almost no difference between the curves. Similarly, there is little difference between the two curves in the super-Ohmic regime in Fig.~\ref{fig8}.  

\subsection{\label{sec:level33} Measurement Interval Dependence}

\begin{figure}[]
\begin{center}
  \includegraphics[scale=0.5]{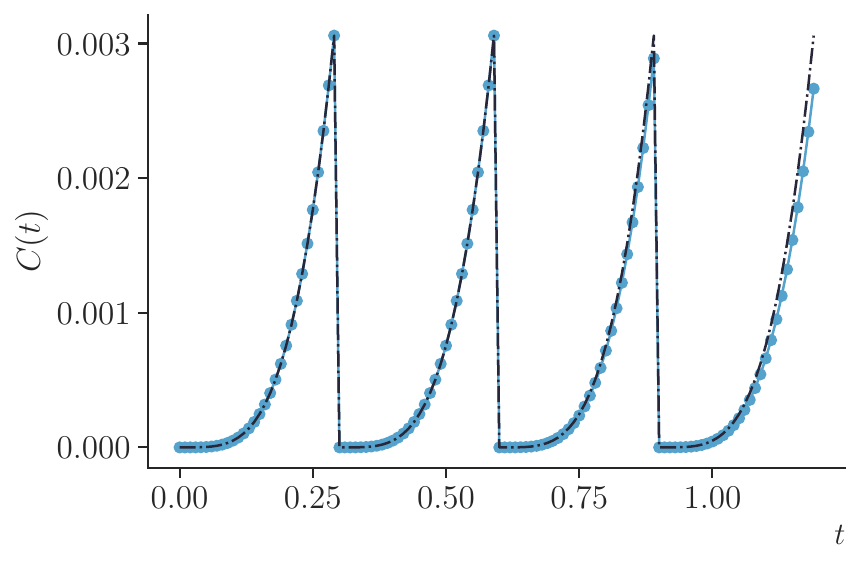}\\
  \caption{Here we are using the Ohmicity parameter $ s = 0.1$ with $G = 1$, and the measurement interval $\tau = 0.3$.} \label{fig9}
  \end{center}
\end{figure}

\begin{figure}[]
\begin{center}
  \includegraphics[scale=0.5]{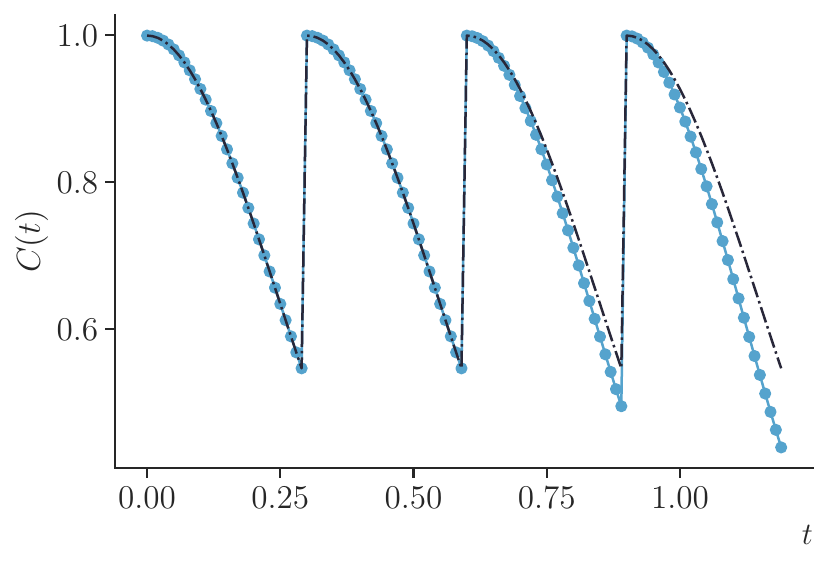}\\
  \caption{Here we are using the Ohmicity parameter $s = 0.1$ with $G = 1$, and the measurement interval $\tau = 0.3$. Note the different state being prepared repeatedly.}\label{fig10}
  \end{center}
\end{figure}

In the previous sections, we looked at entanglement dynamics for different values of the Ohmicity parameter, $s$, while keeping the coupling strength, $G$, and the measurement interval, $\tau$, fixed. For both cases, product and maximally entangled state preparation, we observe significant accumulation of the phase factors, $\gamma_{ij}, \mu_{ij}, \epsilon_{ij}$ and $\sigma_{ij}$ which emerge from $\text{T}_{\text{Env}}$. These phase factors depend on the total time elapsed $t$ and the measurement interval $\tau$. The QZE and QAZE emerge from the dependence on measurement intervals $\tau$. This is the dependence we aim to explore in this section. To this end, we have plotted the entanglement dynamics for the scenario where the product state is repeatedly prepared but with a smaller measurement interval than before (we now set $\tau = 0.3$). Results are illustrated in Fig.~\ref{fig9}. Similarly, we plot the entanglement dynamics with the repeated preparation of the entangled Bell state in Fig.~\ref{fig10}. 
These figures, when compared with Fig.~\ref{fig1} and Fig.~\ref{fig5} respectively, show that with a smaller measurement interval, there is little difference due to the changing environment state. This is expected as the rapid repeated measurement does not allow the influence of the system-environment correlations to manifest themselves. 

Here here onward, we repeatedly prepare the product state defined in Eq.~\eqref{eq:productstate}. For comparison between different Ohmicity parameters, we use the ratio $C_{\text{max}}/C_{\text{reset}}$. $C_{\text{max}}$ is the maximum concurrence (with respect to time) achieved after $N$ measurements with the changing environment state, and $C_{\text{reset}}$ is the maximum concurrence generated in the scenario where the environment is reset. In the following results, we plot this ratio against measurement intervals that span from the Zeno to the anti-Zeno regime for different numbers of measurements.
\begin{figure}[]
\begin{center}
  \includegraphics[scale=0.55]{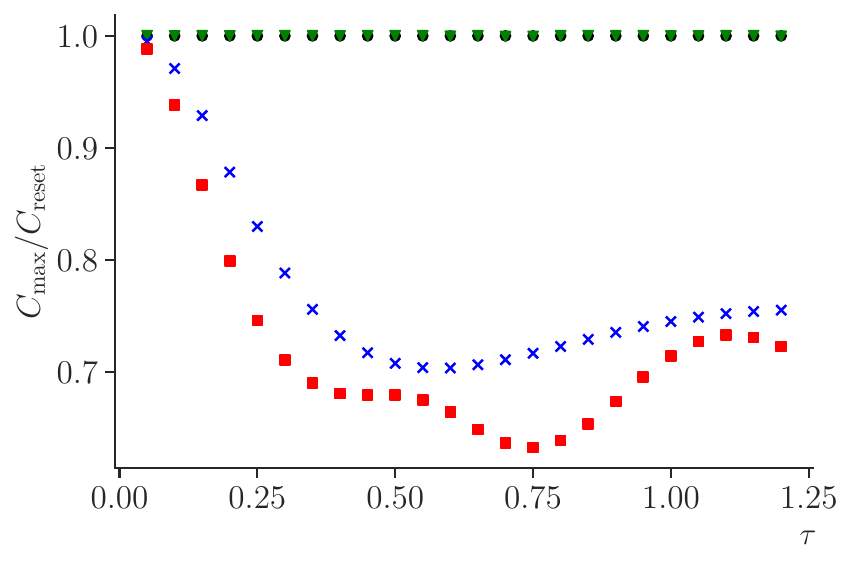}\\
  \caption{This figure shows the effect of the measurement interval on concurrence generation in the sub-Ohmic regime with $s=0.5$. The black circles are obtained when the environment is repeatedly reset. In contrast, the green upside-down triangles correspond to one measurement, the blue crosses to two measurements, and the red squares to three measurements (we take the changing environment state into account).}\label{peak0.5}
  \end{center} 
\end{figure}
In Fig.~\ref{peak0.5}, we have plotted our results for the sub-Ohmic regime. Here, we can see that the results for one measurement are very close to those where the environment reset. However, with an increase in the number of measurements, we see a decrease in the concurrence ratios. This is expected as the changing environment state plays a bigger role with an increasing number of measurements. 
\begin{figure}[]
\begin{center}
  \includegraphics[scale=0.55]{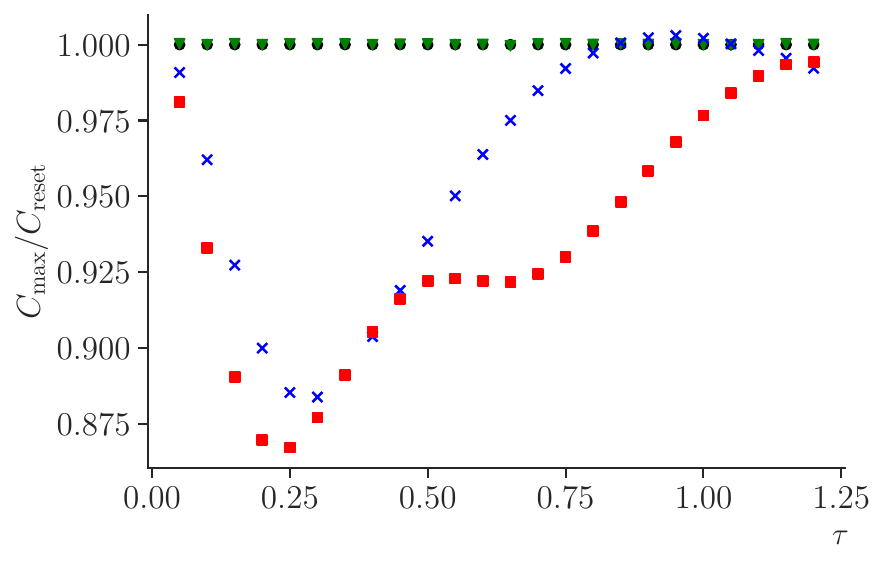}\\
  \caption{This is the same as Fig.~\ref{peak0.5}, except that we now study the Ohmic regime ($s = 1$).}\label{peak1}
  \end{center}
\end{figure}
A similar trend in the Ohmic regime can be seen in Fig.~\ref{peak1}. However, here the role of the changing environment state, while noticeable, is considerably less pronounced.

\section{\label{sec:conc}Conclusion}
In this study, we explore the entanglement dynamics of a two-qubit system interacting with a harmonic oscillator environment, with repeated projective measurements performed on the system. To this end, we calculated the post-measurement system states for two cases. In one of the cases, the environment was reset to the thermal equilibrium state every time a measurement was made. The second case was where the environment did not remain in the thermal equilibrium state as measurements on the system were made. In fact, the system and the environment correlate as a consequence of the system-environment interaction. As a result, any measurement performed on the system affects the environment. Consequently, repeated measurements performed on the system keep changing the environment state. We simulated and compared the entanglement evolution for the scenario where the environment is repeatedly reset to the scenario where the aforementioned environment evolution is considered. We showed that the entanglement dynamics can be significantly different. By altering the measurement interval, we also investigated how the system dynamics are gradually frozen as we reduce the measurement interval. Our work is the first step towards understanding the role of system-environment correlations in studies of the quantum Zeno and anti-Zeno effects in multiqubit systems.

 \begin{acknowledgments}
 We want to thank Salman for his insightful questions and infectious curiosity. It would be remiss of us to forget Nabeel's help during the initial phases of this project. Sometimes, the needs of one's heart require more attention than those of one's mind. We are forever indebted to  Hassan, Maryam, Hafsa, Alina, Azka, Bismah, and Wahaj for providing the required care. 

\end{acknowledgments}

\appendix

\section{ Unitary Time Evolution Operator} \label{appendix}
Here, we present a derivation of the unitary time evolution operator, $U(t)$, using the total Hamiltonian  given by 
\begin{equation}
    H=H_S+H_E+H_{I}, \nonumber
\end{equation} where
\begin{equation} \nonumber
    H_S=\frac{\omega_0}{2}(\sigma_z^{ (1)}+\sigma_z^{(2)}),
\end{equation}
\begin{equation} \nonumber
   H_E=\sum_k \omega_k b_k^\dagger b_k,
\end{equation} and 
\begin{equation}\nonumber
    H_{I}= (\sigma_z^{ (1)}+\sigma_z^{(2)})\sum_r(g_r^* b_r+g_r b_r^\dagger).
\end{equation}
Let us first transform to the interaction picture. That is, with $U_0(t)= e^{-i(H_E+H_S)t}$, we find $H_{I}(t)$ to be
\begin{align*}
    H_{I}(t) &=U_0(t)^\dagger H_{I} U_0(t) \\
    &=(\sigma_z^{(1)}+\sigma_z^{(2)})\sum_r(g_r^* b_r e^{-i\omega_rt}+g_r b_r^\dagger e^{i\omega_r t}) \nonumber.
\end{align*}
We now move on to find the time evolution operator in the interaction picture, $U_I(t)$, using the Magnus expansion \cite{magnus,PhysRevA.104.042205,Blanes_2010}, 
\begin{equation}
    U_I(t)=e^{\sum^\infty _{i=1} A_i(t)},
\end{equation}
where 
\begin{align}\nonumber
    &A_1=-i\int_0^t H_I(t_1)dt_1,\\\nonumber
    &A_2=-\frac{1}{2} \int_0^t dt_1 \nonumber\int_0^{t_1} dt_2 \, [H_I(t_1),H_I(t_2)].
\end{align}
It can be shown that 
\begin{align*}
    &A_1= \frac{1}{2}(\sigma_z^{(1)}+\sigma_z^{(2)}) \sum_r (\alpha_r(t)b_r^\dagger-\alpha^*_r(t)b_r), \\
    &A_2= -\frac{i}{2}(\mathds{1}+\sigma_z^{(1)} \sigma_z^{(2)})\Delta(t),
\end{align*}
where 
\begin{align*}
    \Delta (t)=\sum _r \frac{4|g_r|^2}{\omega_r^2}[\sin(\omega_r t)-\omega_r t]
\end{align*}
and 
\begin{equation*}
\alpha_r(t)=\frac{2 g_r\left(1-e^{i \omega_r t}\right)}{\omega_r}.
\end{equation*}
One can also check that all subsequent terms, $A_3, A_4,...$, are zero. 
Consequently, the total unitary time evolution operator takes the form
\begin{equation}\label{unitary}
    U(t)=U_0 U_I(t)
    =e^{-i\frac{ \omega_o}{2}(\sigma_z^{(1)}+\sigma_z^{(2)})t} e^{- i H_E t} U_I(t).
\end{equation}

\section{Reset The Environment} \label{appendb}
We now calculate the density matrix in the presence of measurements, but where the environment is repeatedly assumed to reset to the thermal equilibrium state. We start with
\begin{equation}
\rho_S(t)=\text{Tr}_E\left[U(t) \rho(0)U^{\dagger}(t)\right].
\end{equation}
We want to express this density matrix in the eigenbasis of $\sigma_z^{(1)}$ and $\sigma_z^{(2)}$ given by $\sigma_z^{(1)}|k, l\rangle=k|k, l\rangle$ and $\sigma_z^{(2)}|k, l\rangle=l|k, l\rangle$. To do this we introduce the operator $P_{kl,k'l'} = \ket{k,l}\bra{k',l'} $. We can then write
\begin{align}\nonumber
[\rho_S(t)]_{k'l',kl} &=\operatorname{Tr}_{S, E}[U(t) \rho(0) U^{\dagger}(t) P_{kl,k'l'}] \\
& =\operatorname{Tr}_{S, E}[\rho(0) P_{kl,k'l'}(t)]\nonumber.
\end{align}
Here $P_{kl,k'l'}(t)= U^{\dagger}(t) P_{kl,k'l'} U(t)$. We now use the form of the unitary time evolution operator we found previously in Eq.~\eqref{unitary} to show that $P_{kl,k'l'}(t)= e^{i \frac{\omega_0}{2}\left(k+l-k^{\prime}-l^{\prime}\right) t} e^{i \frac{\Delta(t)}{2}\left(k l-k^{\prime} l^{\prime}\right)} e^{-R_{kl,k'l'}(t)} P_{kl,k'l'}$. Note that  
\begin{equation*}
\begin{aligned}
R_{kl,k'l'}(t) & = \frac{1}{2}\left(k+l-k^{\prime}-l^{\prime}\right) \sum_r\left[\alpha_r(t) b_r^{\dagger}-\alpha_r^*(t) b_r\right]. \\
\end{aligned}
\end{equation*}
Assembling all these pieces, we can write the density operator of the system as
\begin{equation}
\begin{aligned}
\left[\rho_S(t)\right]_{k'l',kl}&= e^{-i \frac{\omega_0}{2}\left(k^{\prime}+l^{\prime}-k-l\right) t} e^{-i \frac{\Delta(t)}{2}\left(k^{\prime} l^{\prime}-k l\right)} \\
 &\times  \operatorname{Tr}_{S, E}\left[\rho(0) e^{-R_{kl,k'l'}(t)} P_{kl,k'l'}\right].
\end{aligned}
\end{equation}
As we have discussed before, we are interested in the case where $\rho(0)=\rho_S (0) \otimes \rho_E (0)$, where $\rho_E (0)=\rho_E=e^{-\beta H_E}/Z_E$ (we assume the environment remains unchanged after each measurement), with $\beta$ the inverse temperature and $Z_E=\operatorname{Tr}_E\left[e^{-\beta H_E}\right]$. This leads to   
\begin{align} \label{chulo}
{\left[\rho_S(t)\right]_{k'l',kl} }&= \left[\rho_S(0)\right]_{k'l',kl}  e^{-i \frac{\omega_0}{2}\left(k^{\prime}+l^{\prime}-k-l\right) t} \\
&\times e^{-i \frac{\Delta(t)}{2}\left(k^{\prime} l^{\prime}-k l\right)} \operatorname{Tr}_E\left[\rho_E (0) e^{-R_{kl,k'l'}(t)}\right]. \nonumber
\end{align}
Notice that $\operatorname{Tr}_E\left[\rho_E  e^{-R_{kl,k'l'}(t)}\right]=\left\langle e^{-R_{kl,k'l'}(t)}\right\rangle$. This can be simplified as 
\begin{align*}
\left\langle e^{-R_{kl,k'l'}(t)}\right\rangle= e^ {-\frac{1}{4}\left(k+l-k^{\prime}-l^{\prime}\right)^2 \gamma(t)},
\end{align*}
where
\begin{equation*}
    \gamma(t)=\sum_r \frac{4\left|g_r\right|^2}{\omega_r^2}\left[1-\cos \left(\omega_r t\right)\right] \operatorname{\coth}\left(\frac{\beta \omega_r}{2}\right).
\end{equation*}
Here, we have used the form of the Bose-Einstein distribution. We converted $\Delta(t)$ and $\gamma(t)$ into integrals using the spectral density function to perform numerical calculations. Finally, we can write the system density operator as 
\begin{equation}
\begin{aligned}
\left[\rho_S(t)\right]_{k'l',kl}&=\left[\rho_S(0)\right]_{k'l',kl} e^{-i \frac{\omega_0}{2}\left(k^{\prime}+l^{\prime}-k-l\right) t}\\ 
& \times e^{-i \frac{\Delta(t)}{2}\left(k^{\prime} l^{\prime}-k l\right)} e^{-\frac{1}{4}\left(k+l-k^{\prime}-l^{\prime}\right)^2 \gamma(t)}
\end{aligned}
\end{equation}
The system density operator presented above is for the case where whenever a measurement is made, the system collapses to a projection state, and the environment is reset to its thermal equilibrium state. However, this is not the scenario we are interested in.
We want to study the more realistic scenario (especially in the strong system-environment coupling regime) in which the environment changes over time with measurements.

\section{Keeping Track of The Changing Environment State} \label{appendc}

To start the calculation for the case where we keep track of the changing environment state with the measurements performed, we start with the fact that the total system-environment density matrix  after $N - 1$ measurements is
\begin{widetext}
\begin{align} 
    \rho \big ((N-1)\tau \leq t < N\tau  \big ) &= \frac{1}{Z_{N-1}} U \big(t-\tau(N-1)\big) \big[P_\psi U (\tau) \big]^{N-1}\rho(0) \big [U^\dagger (\tau) P_\psi \big]^{N-1} U^\dagger \big(t-\tau(N-1)\big), \\
    \label{eq:5}
    Z_{N-1} &= \text{Tr}_{S,E}\bigg\{ \big[P_\psi U (\tau) \big]^{N-1}\rho(0) \big [U^\dagger (\tau) P_\psi \big]^{N-1} U^\dagger \big(t-\tau(N-1)\big)  \bigg\}.
\end{align}
\end{widetext}
$N$ is an integer not less than one. Here, $U(t)$ is the total system-environment time-evolution operator we derived previously, and $P_\psi$ is the projection operator onto the state $\ket{\psi}$. We also define $t'=t-(N - 1)\tau$. Now, tracing out the environment, we obtain 
\begin{align}
\rho_S(t)=  &\frac{1}{Z_{N-1}}  \text{Tr}_{E} \big \{ U \big(t'\big) \big[P_\psi U (\tau) \big]^{N-1}\\
&\times \rho(0)\big [U^\dagger (\tau) P_\psi \big]^{N-1} U^\dagger \big(t'\big) \big \}.  \nonumber
\end{align}
As we have done before, we would now like to extract the elements of the system density matrix by
\begin{align} \label{wow}
[\rho_S(t)]_{k'l',kl}= &\frac{1}{Z_{N-1}} \text{Tr}_{S,E}\big \{  \big[P_\psi U (\tau) \big]^{N-1}\rho(0) \\
         & \times \big [U^\dagger (\tau) P_\psi \big]^{N-1}   P_{kl,k'l'}(t') \big \}. \nonumber
\end{align}
Now, we assume that the initial density matrix is a product state initially, that is,
$\rho(0)=P_\psi \otimes \rho_E(0)$, where  $\rho_E(0)$ is environment thermal equilibrium state. This does not mean the environment state will always be in this thermal equilibrium state. Now, by an inductive proof, one can show that Eq.~\eqref{wow} becomes
\begin{widetext}
    \begin{align*}
{[\rho_S(t)]}_{k'l',kl}=  \frac{e^{\frac{i}{2}\Delta(t')(kl-k'l')}}{Z_{N-1}} \text{Tr}_{S,E}\big\{ \ket{\psi}  \big[\bra{\psi} U (\tau)\ket{\psi} \big ]^{N-1}\rho_{E} (0)\big[\bra{\psi} U^\dagger (\tau) \ket{\psi} \big ]^{N-1}\bra{\psi}P_{kl,k'l'} e^{-R_{kl,k'l'}(t')} \big \}.
    \end{align*}
\end{widetext}
We can express this as a product of a trace over the environment and a trace over the system,
\begin{equation} 
[\rho_S(t)]_{k'l',kl}=\frac{e^{\frac{i}{2}\Delta(t')(kl-k'l')}}{Z_{N-1}} \text{T}_{\text{env}}\text{T}_{\text{sys}},
\end{equation}
where $\text{T}_{\text{env}}$ takes the form,
\begin{align*}
    \text{T}_{\text{env}} =& \text{Tr}_{E}\big \{  \big[\bra{\psi} U (\tau) \ket{\psi} \big ]^{N-1}\rho_{E}(0) \\
    & \times \big[\bra{\psi} U^\dagger (\tau) \ket{\psi} \big ]^{N-1}  e^{-R_{kl,k'l'}(t')} \big \}. \\
\end{align*}
On the other hand, $\text{T}_{\text{sys}}$ takes the form,
\begin{align*}
    \text{T}_{\text{sys}} =& \text{Tr}_{S} \big \{P_\psi P_{kl,k'l'}\}.
\end{align*}
To make further progress, we note that
\begin{equation}
    \bra{\psi} U (\tau) \ket{\psi}= e^{-iH_E \tau } X(\tau),
\end{equation} 
where   $X(\tau)= \sum_{m,n} \left|\ip{m,n}{\psi}\right|^2 e^{-\frac{i}{2}(1+mn)\Delta(\tau)+R^{m,n}(\tau)}$ and  $R^{m,n}=\frac{1}{2}(m+n)\sum_r [\alpha_r(\tau) b_r^\dagger-\alpha^*_r(\tau) b_r]$. Note that this quantity is comprised of environment operators only. 
Consequently, it follows,
\begin{widetext}
\begin{equation} \label{trace}
    [\rho_S(t)]_{k'l',kl}= \bra{\psi}\ket{k,l}\bra{k',l'} \ket{\psi} \frac{e^{\frac{i}{2}\Delta(t')(kl-k'l')}}{Z_{N-1}} \text{Tr}_{E}\big \{ \rho_{E}(0) \big[X^\dagger(\tau) e^{iH_E \tau } ]^{N-1}  e^{-R_{kl,k'l'}(t')} \big[e^{-iH_E \tau } X(\tau) ]^{N-1} \big \}.
\end{equation}
\end{widetext}

To proceed, we now deal with terms inside the trace in \eqref{trace}. We start with 
$\big[X^\dagger(\tau) e^{iH_E \tau } ]^{N-1}$. This can be shown to equal,
\begin{equation} \label{kia baat}
    e^{-iH_E\tau}X^\dagger_1(\tau) X^
    \dagger_2(\tau)\cdots X^\dagger_{N-1} e^{iN H_E \tau},
\end{equation}
where 
\begin{align*} 
X^\dagger_{p}(\tau)&= e^{pi H_E \tau}X^\dagger e^{-pi H_E \tau} \\ &= \sum_{m,n} |\bra{m,n}\ket{\psi}|^2 e^{\frac{-i}{2}(1+mn)\Delta(\tau)+R^{m,n}_p (\tau)},
\end{align*}
\begin{align*}    
R^{m,n}_p (\tau)&=\frac{1}{2}(m+n)\sum_r \big[\alpha_{r, p}(\tau) b_r^\dagger -\alpha_{r, p}^*(\tau) b_r\big], 
\end{align*} and 
\begin{align*}
    \alpha_{r, p}(\tau)=\frac{2 g_r e^{ip \omega_r \tau}\left(1-e^{i \omega_r \tau}\right)}{\omega_r}. 
\end{align*}
Similarly, we can deal with $ \big[e^{-iH_E \tau } X(\tau) ]^{N-1}$ to get,
\begin{equation}
    e^{-iNH_E\tau} X_{N-1}(\tau) X
   _{N-2}(\tau)\cdots X_{1} e^{i H_E \tau}.
\end{equation}
We now use the two previous results and the cyclic invariance property of the trace to get, 
\begin{widetext}
    \begin{align} \label{kyun}
     [\rho_S(t)]_{k'l',kl}= & \psi^*_{k,l} 
     \psi_{k',l'}\frac{e^{\frac{i}{2}\Delta(t')(kl-k'l')}}{Z_{N-1}}\\ & \times \text{Tr}_{E}\big \{ \rho_{E}(0)  X^\dagger_1(\tau) \cdots X^\dagger_{N-1} e^{iN H_E \tau}    e^{-R_{kl,k'l'}(t')} e^{-iNH_E\tau} X_{N-1}(\tau) \cdots X_{1} \}, \nonumber
\end{align}
\end{widetext}
where for the sake of brevity $\bra{\psi}\ket{k,l} \bra{k',l'} \ket{\psi}=  \psi^*_{k,l} \psi_{k',l'}$. 

We also combine $e^{iN H_E \tau}    e^{-R_{kl,k'l'}(t')} e^{-iNH_E\tau}$ into a single exponential to get 
\begin{equation*}
    e^{iN H_E \tau}    e^{-R_{kl,k'l'}(t')} e^{-iNH_E\tau}= e^{ -\Tilde{R}_{kl,k'l',N}(t',\tau)},
\end{equation*}
where $\Tilde{R}_{kl,k'l',N}(t',\tau)$ is 
 \begin{equation*}
     \frac{1}{2}(k+l-k'-l')\sum_r (\Tilde{\alpha}_{r,N}(t'\tau)b_r^\dagger-\Tilde{\alpha}^*_{r,N}(t',\tau)b_r)
 \end{equation*}
and 
 \begin{align*}
 \Tilde{\alpha}_{r,N}(t',\tau)= e^{i\omega_r N\tau}\alpha_r(t').
 \end{align*}
To proceed, we now attempt to combine $X^\dagger_1(\tau) \cdots X^\dagger_{N-1}(\tau) $ into a single exponential.  Combining each of these gives rise to phase factors of the form, 
\begin{align*}
\frac{[R^{x,y}_p(\tau), R^{z,h}_{p'}(\tau)]}{2}=\frac{i(x+y)(z+h)\mu_{pp'}(\tau)}{2}.
\end{align*}
We have that $\mu_{pp'}(\tau)= \sum_{r}\frac{4|g_r|^2}{\omega_r} \sin\big[(p-p')\omega_r \tau\big]\big[1-\cos(\omega_r \tau)\big]$.
Notice that $\mu_{pp'}=-\mu_{p'p}$ and $\mu_{pp}=0$. Luckily, phase factors apart from the one above disappear because $\mu_{pp'}$ is a c-number. 
 Consequently, we have
 \begin{widetext}
      \begin{align}
     X^\dagger_1(\tau) \cdots X^\dagger_{N-1}(\tau)=& \sum_{m,n} \cdots \sum_{u,v} e^{\frac{i}{2}\Delta(\tau)(mn+..... + uv)} \psi^2_{mn}\cdots \psi^2 _{uv} e^{-R^{m,n}_1..... -R^{u,v}_{N-1}} \\ & \times
     e^{\frac{i}{2}(p+q)(m+n) \mu_{12}+ \frac{i}{2}(s+t)[(m+n)\mu_{13}+(p+q)\mu_{23}]+....+\frac{i}{2}(u+v)[(m+n)\mu_{1(N-1)}+....+ (e+f)\mu_{(N-2)(N-1)}] }\nonumber,
 \nonumber
    \end{align}
 \end{widetext}
where we have defined terms like $\left|\bra{m,n}\ket{\psi}\right|^2$ as $\psi^2_{mn}$. This expression needs some explanation.  $\mu_{p p'}$ is a factor picked up for each term combined in $X^\dagger_1(\tau) \cdots X^\dagger_{N-1}$ and $p< p'$ always. Each combined term has associated indices that also appear in the phase factors. For example, for  $X^\dagger_1(\tau)$ and $X^\dagger_{N-1}$, we have the indices $(m,n)$ and $(u,v)$. Similarly, we have

\begin{widetext}
 \begin{align}
      X_{N-1}(\tau) \cdots X_{1}(\tau)  &=\sum_{u',v'} \cdots \sum_{m',n'} e^{-\frac{i}{2}\Delta(\tau)(u'v'+.... + m'n')} \psi^2_{u'v'}\cdots \psi^2 _{m'n'} e^{R^{u',v'}_{N-1}+.... +R^{m',n'}_{1}}\\ & \times
      e^{-\frac{i}{2}(u'+v')(b'+c')\mu_{(N-2)(N-1)}\cdots -\frac{i}{2}(m'+n')[(u'+v') \mu_{1(N-1)}+.... + (p'+q')\mu_{1 2}]}.\nonumber
 \end{align}
\end{widetext}
Here, we have used primed indices when combining terms. The summations and phase factors can be taken out of the trace in Eq.~\eqref{kyun}; that leaves us with 

\begin{equation*}
    \text{Tr}_{E} \big \{ \rho_{E}(0) e^{-R^{m,n}_1....-R^{u,v}_{N-1}}  e^{\Tilde{R}_{kl,k'l',N}}    e^{R^{u',v'}_{N-1}+.... +R^{m',n'}_{1}}\big\}.
\end{equation*}
We combine the first exponent in the trace with the exponent term in the center. We will need to deal with phase factors of the form, 

\begin{align*}
    \frac{1}{2}[-R_p^{w, x},- \Tilde{R}_{kl, k'l',N}]. 
\end{align*} 
This daunting-looking commutator takes the form,
\begin{widetext}
    \begin{align*}
        (w+x)(k+l-k'-l')\sigma_{p,N}(t',\tau)=& (w+x)(k+l-k'-l')\sum_r \frac{2i|g_r|^2}{w_r^2}\bigg\{-\sin\big[\omega_r(p-N)\tau -\omega_r t'\big] \\ &-\sin\big[\omega_r(p-N+1)\tau \big]+\sin\big[\omega_r(p-N)\tau\big]+\sin\big[\omega_r(p-N+1)\tau -\omega_r t'\big]\bigg\}.
    \end{align*}
\end{widetext}
That leaves us with the following scenario inside the trace,
\begin{equation*}
    \text{Tr}_E \big \{ \rho_{E}(0) e^{-R^{m,n}_1....-R^{u,v}_{N-1}+\Tilde{R}_{kl,k'l',N}}  e^{R^{u',v'}_{N-1}+.... +R^{m',n'}_{1}}\big\}.
\end{equation*}
Note that we have not written down the phase factors for now. Finally, we combine the exponents in this expression as well. We must use the previously defined $\mu(\tau)$ and $\sigma(t',\tau)$. The form of the trace now becomes,
 \begin{equation*}
    \text{Tr}_E\bigg\{\rho_{E}(0)e^{-R_1^{m,n}....-R_{N-1}^{u,v}+ \tilde{R}_{kl,k'l',N}+R_{N-1}^{u',v'}+....+R_1^{m',n'}}  \bigg\}.
\end{equation*}
Now, we group the primed and unprimed indices,

\begin{equation*}
    e^{(-R_1^{m,n}+R_1^{m',n'})+(-R_2^{p,q}+R_2^{p',q'}).....(-R_{N-1}^{u,v}+R_{N-1}^{u',v'})-\tilde{R}_{kl,k'l',N}}   
\end{equation*}
and define, for example, $R_1^{m',n'}-R_1^{m,n}=\frac{1}{2}[m'+n'-m-n]R_1$ as $R_1^{m',n',m,n}$. This gives us 

\begin{equation*}
    \text{Tr}_{E}\bigg\{ \rho_{E}(0)e^{R_1^{m',n',m,n}+R_2^{p',q',p,q}+....R_{N-1}^{u',v',u,v}-\tilde{R}_{kl,k'l',N}}\bigg\}.
\end{equation*}
For our purposes, $\rho_{E}(0)= \frac{e^{-\beta H_E}}{Z_E}$ and $Z_E=\operatorname{Tr}_E\left[e^{-\beta H_E}\right] $. This allows us to carry out steps similar to the ones after Eq.~\eqref{chulo}. This gives us four terms $\big<R_{p}R_{p'}+R_{p'}R_{p}\big>$, $\big<R_{p}^2\big>$, $\big<\tilde{R}^{2}\big>$, and $\big<\tilde{R} R_{p}+R_{p}\tilde{R}\big>$, ignoring some indices. Note that all these terms have a sum over all modes. The first term $\big<R_{p}R_{p'}+R_{p'}R_{p}\big>$ is defined as $-4\gamma_{pp'}(\tau)$. We have that

\begin{widetext}
    \begin{equation*}
    \gamma_{pp'} (\tau)= \sum_r \frac{4|g_{r}|^2}{w_r^2}\bigg\{[1-\cos(\omega_r\tau)]\cos[\omega_r\tau(p-p')]\coth\bigg[\frac{\beta\omega_r}{2}\bigg ]\bigg\}.
\end{equation*}
\end{widetext}

Observe that $ \gamma_{pp'}=\gamma_{p'p}$. Next, we obtain $\big<R_{p}^2\big>$ as,
\begin{equation*}
\sum_r |\alpha_r^2|\coth\bigg\{\frac{\beta\omega_r}{2}\bigg\}\equiv2\gamma(\tau).
\end{equation*}
Now, we have $\big<\tilde{R}^{2}\big>$ which becomes,
\begin{equation*}
    - \sum_r| \tilde{\alpha}_{r,N}|\coth\bigg(\frac{\beta\omega_k}{2}\bigg)\equiv -2\gamma (t').
\end{equation*}
The final term in our expression, $\big<\tilde{R}_{N}R_{p}+R_{p}\tilde{R}_{N}\big>$ is obtained as
\begin{align*}
&-8\sum_{r}\frac{|g_r|^2}{\omega_{r^2}}\coth\bigg(\frac{\beta\omega_r}{2}\bigg)\bigg\{\cos[(p-N)\omega_{r} \tau]-\\
 & \cos[(p-N+1)\omega_{r} \tau]-\cos[(p-N)\omega_{r} \tau-\omega_{r}t']\\
& +\cos[(p-N+1)\omega_{r} \tau-\omega_r t']\bigg\}.
\end{align*}
We define this term as $-2\epsilon_{p,N}(t',\tau)$.
The four terms above can be made into integrals using the spectral density, as was done previously.

Finally, we have all the ingredients to write the general expression of the density matrix of the two qubits at any time after the $(N-1)$\textsuperscript{th}measurement and  before the $N$\textsuperscript{th} measurement: 

\begin{widetext}
    \begin{align} \label{eq:FinalExpanded}
      [\rho_S(t)]_{k'l',k,l} =& \frac{1}{Z_{N-1}}\psi_{k',l'} \psi^*_{k,l}e^{\frac{i}{2}\Delta(t')(kl-k'l')}\sum_{mn}....\sum_{uv}\sum_{u'v'}....\sum_{m'n'}\psi_{mn}^2 .... \psi_{uv}^2 \psi_{u'v'}^{2}....\psi_{m'n'}^{2}\\ \nonumber
      & \times e^{i\mu_{12} (\tau)[(m+n)(p+q)-(m'+n')(p'+q')+(m'+n')(p+q)-(m+n)(p'+q')]}....\\ \nonumber
      & \times e^{i\mu_{(N-2)(N-1) }(\tau)[(b+c)(u+v)-(b'+c')(u'+v')+(b'+c')(u+v)-(b+c)(u'+v')]}\\ \nonumber
      & \times e^{-\frac{1}{4}[(m'+n'-m-n)^{2}....(u'+v'-u-v)^{2}]\gamma(\tau)}e^{\frac{i}{2}\Delta(\tau)(mn-m'n'+....uv-u'v')}\\ \nonumber
     & \times e^{-\frac{1}{4}[(m'+n'-m-n)(p'+q'-p-q)\gamma_{12}(\tau)]}....e^{-\frac{1}{4}[(b'+c'-b-c)(u'+v'-u-v)]\gamma_{(N-2)(N-1)}(\tau)}\\ \nonumber
     & \times e^{- \frac{1}{4}[(k+l-k'-l')^{2}]\gamma(t')}e^{-\frac{1}{4}(k+l-k'-l')[(m'+n'-m-n)\epsilon_{1,N}(t',\tau)+....(u'+v'-u-v)\epsilon_{N-1,N}(t',\tau)]}\\ \nonumber
     & \times e^{\frac{1}{2}(k+l-k'-l')[(m'+n'+m+n)\sigma_{1,N}(t',\tau)+....(u'+v'+u+v)\sigma_{N-1,N}(t',\tau)]},\\ \nonumber
    \end{align}
\end{widetext}
\begin{widetext}
where
    \begin{align} 
Z_{N-1}=&\sum_{mn}....\sum_{uv}\sum_{u'v'}....\sum_{m'n'}\psi_{mn}^2 .... \psi_{uv}^2 \psi_{u'v'}^{2}....\psi_{m'n'}^{2}\\ \nonumber
      & \times e^{i\mu_{12} (\tau)[(m+n)(p+q)-(m'+n')(p'+q')+(m'+n')(p+q)-(m+n)(p'+q')]}....\\ \nonumber
      & \times e^{i\mu_{(N-2)(N-1) }(\tau)[(b+c)(u+v)-(b'+c')(u'+v')+(b'+c')(u+v)-(b+c)(u'+v')]}\\ \nonumber
      & \times e^{-\frac{1}{4}[(m'+n'-m-n)^{2}....(u'+v'-u-v)^{2}]\gamma(\tau)}e^{\frac{i}{2}\Delta(\tau)(mn-m'n'+....uv-u'v')}\\ \nonumber
     & \times e^{-\frac{1}{4}[(m'+n'-m-n)(p'+q'-p-q)\gamma_{12}(\tau)]}....e^{-\frac{1}{4}[(b'+c'-b-c)(u'+v'-u-v)]\gamma_{(N-2)(N-1)}(\tau)}.\\ \nonumber
    \end{align}
\end{widetext}
Let us examine closely the two final expressions. Remember that $(b,c)$ and $(u,v)$ correspond to the $(N-2)$\textsuperscript{th} and the $(N-1)$\textsuperscript{th} measurement  respectively, and $(m,n)$  and $(p,q)$ correspond to the first and second measurements respectively. We have a sum over the indices $(m,n)$ and $(p,q)$ up to $(b,c)$ and $(u,v)$. Similarly, there is a sum over the same indices but primed. There are projection factors like $\psi^2_{mn}$ for all primed and non-primed indices. In terms like $\mu_{jk}$, the subscript is determined by forming pairs of numbers from 1 to  $N-1$ . This pair will be of the form $(j, k)$, where $j<k$. What about the factors multiplying the $\mu_{jk}$ terms in the exponent? Consider $\mu_ {12}$. The factors that get multiplied are primed and non-primed indices associated with ``1" and ``2" in the subscript, but these factors follow the pattern in the expressions above. This is how the $\gamma_{jk}$ terms also operate. For $\epsilon_{j N}$ and $\sigma_{jN}$, the second subscript, $N$, is fixed, but $j$ can go from 1 to $N-1$ . The factors that tag along with these terms are again comprised of their respective indices according to the pattern above. Essentially, $\gamma_{jk}$, $\epsilon_{jk}$, $\sigma_{jk}$, and $\mu_{jk}$ allow us to take into account the impact of measurement on decoherence of the system as the environment keeps evolving due to measurement. The $\Delta(t')$ and $\Delta(\tau)$ terms quantify the indirect interaction between the qubits due to the common environment. Finally, the normalization factor of $Z_{N-1}$ is in the denominator. It has a sum over the same indices as the numerator and has the same terms as the numerator, except that the terms that depend on $t'$ are absent.


\begin{thebibliography}{35}%
\makeatletter
\providecommand \@ifxundefined [1]{%
 \@ifx{#1\undefined}
}%
\providecommand \@ifnum [1]{%
 \ifnum #1\expandafter \@firstoftwo
 \else \expandafter \@secondoftwo
 \fi
}%
\providecommand \@ifx [1]{%
 \ifx #1\expandafter \@firstoftwo
 \else \expandafter \@secondoftwo
 \fi
}%
\providecommand \natexlab [1]{#1}%
\providecommand \enquote  [1]{``#1''}%
\providecommand \bibnamefont  [1]{#1}%
\providecommand \bibfnamefont [1]{#1}%
\providecommand \citenamefont [1]{#1}%
\providecommand \href@noop [0]{\@secondoftwo}%
\providecommand \href [0]{\begingroup \@sanitize@url \@href}%
\providecommand \@href[1]{\@@startlink{#1}\@@href}%
\providecommand \@@href[1]{\endgroup#1\@@endlink}%
\providecommand \@sanitize@url [0]{\catcode `\\12\catcode `\$12\catcode
  `\&12\catcode `\#12\catcode `\^12\catcode `\_12\catcode `\%12\relax}%
\providecommand \@@startlink[1]{}%
\providecommand \@@endlink[0]{}%
\providecommand \url  [0]{\begingroup\@sanitize@url \@url }%
\providecommand \@url [1]{\endgroup\@href {#1}{\urlprefix }}%
\providecommand \urlprefix  [0]{URL }%
\providecommand \Eprint [0]{\href }%
\providecommand \doibase [0]{https://doi.org/}%
\providecommand \selectlanguage [0]{\@gobble}%
\providecommand \bibinfo  [0]{\@secondoftwo}%
\providecommand \bibfield  [0]{\@secondoftwo}%
\providecommand \translation [1]{[#1]}%
\providecommand \BibitemOpen [0]{}%
\providecommand \bibitemStop [0]{}%
\providecommand \bibitemNoStop [0]{.\EOS\space}%
\providecommand \EOS [0]{\spacefactor3000\relax}%
\providecommand \BibitemShut  [1]{\csname bibitem#1\endcsname}%
\let\auto@bib@innerbib\@empty
\bibitem [{\citenamefont {Misra}\ and\ \citenamefont
  {Sudarshan}(2008)}]{10.1063/1.523304}%
  \BibitemOpen
  \bibfield  {author} {\bibinfo {author} {\bibfnamefont {B.}~\bibnamefont
  {Misra}}\ and\ \bibinfo {author} {\bibfnamefont {E.~C.~G.}\ \bibnamefont
  {Sudarshan}},\ }\bibfield  {title} {\bibinfo {title} {{The Zeno’s paradox
  in quantum theory}},\ }\href {https://doi.org/10.1063/1.523304} {\bibfield
  {journal} {\bibinfo  {journal} {Journal of Mathematical Physics}\ }\textbf
  {\bibinfo {volume} {18}},\ \bibinfo {pages} {756} (\bibinfo {year}
  {2008})}\BibitemShut {NoStop}%
\bibitem [{\citenamefont {Kwiat}\ \emph {et~al.}(1999)\citenamefont {Kwiat},
  \citenamefont {White}, \citenamefont {Mitchell}, \citenamefont {Nairz},
  \citenamefont {Weihs}, \citenamefont {Weinfurter},\ and\ \citenamefont
  {Zeilinger}}]{kwiat1999high}%
  \BibitemOpen
  \bibfield  {author} {\bibinfo {author} {\bibfnamefont {P.~G.}\ \bibnamefont
  {Kwiat}}, \bibinfo {author} {\bibfnamefont {A.}~\bibnamefont {White}},
  \bibinfo {author} {\bibfnamefont {J.}~\bibnamefont {Mitchell}}, \bibinfo
  {author} {\bibfnamefont {O.}~\bibnamefont {Nairz}}, \bibinfo {author}
  {\bibfnamefont {G.}~\bibnamefont {Weihs}}, \bibinfo {author} {\bibfnamefont
  {H.}~\bibnamefont {Weinfurter}},\ and\ \bibinfo {author} {\bibfnamefont
  {A.}~\bibnamefont {Zeilinger}},\ }\bibfield  {title} {\bibinfo {title}
  {High-efficiency quantum interrogation measurements via the quantum zeno
  effect},\ }\href {https://link.aps.org/doi/10.1103/PhysRevLett.83.4725}
  {\bibfield  {journal} {\bibinfo  {journal} {Physical Review Letters}\
  }\textbf {\bibinfo {volume} {83}},\ \bibinfo {pages} {4725} (\bibinfo {year}
  {1999})}\BibitemShut {NoStop}%
\bibitem [{\citenamefont {Franson}\ \emph {et~al.}(2004)\citenamefont
  {Franson}, \citenamefont {Jacobs},\ and\ \citenamefont
  {Pittman}}]{franson2004quantum}%
  \BibitemOpen
  \bibfield  {author} {\bibinfo {author} {\bibfnamefont {J.~D.}\ \bibnamefont
  {Franson}}, \bibinfo {author} {\bibfnamefont {B.~C.}\ \bibnamefont
  {Jacobs}},\ and\ \bibinfo {author} {\bibfnamefont {T.~B.}\ \bibnamefont
  {Pittman}},\ }\bibfield  {title} {\bibinfo {title} {Quantum computing using
  single photons and the zeno effect},\ }\href
  {https://link.aps.org/doi/10.1103/PhysRevA.70.062302} {\bibfield  {journal}
  {\bibinfo  {journal} {Physical Review A}\ }\textbf {\bibinfo {volume} {70}},\
  \bibinfo {pages} {062302} (\bibinfo {year} {2004})}\BibitemShut {NoStop}%
\bibitem [{\citenamefont {Gordon}\ \emph {et~al.}(2010)\citenamefont {Gordon},
  \citenamefont {Rao},\ and\ \citenamefont {Kurizki}}]{Gordon_2010}%
  \BibitemOpen
  \bibfield  {author} {\bibinfo {author} {\bibfnamefont {G.}~\bibnamefont
  {Gordon}}, \bibinfo {author} {\bibfnamefont {D.~D.~B.}\ \bibnamefont {Rao}},\
  and\ \bibinfo {author} {\bibfnamefont {G.}~\bibnamefont {Kurizki}},\
  }\bibfield  {title} {\bibinfo {title} {Equilibration by quantum
  observation},\ }\href {https://doi.org/10.1088/1367-2630/12/5/053033}
  {\bibfield  {journal} {\bibinfo  {journal} {New Journal of Physics}\ }\textbf
  {\bibinfo {volume} {12}},\ \bibinfo {pages} {053033} (\bibinfo {year}
  {2010})}\BibitemShut {NoStop}%
\bibitem [{\citenamefont {Zhu}\ \emph {et~al.}(2014)\citenamefont {Zhu},
  \citenamefont {Gadway}, \citenamefont {Foss-Feig}, \citenamefont
  {Schachenmayer}, \citenamefont {Wall}, \citenamefont {Hazzard}, \citenamefont
  {Yan}, \citenamefont {Moses}, \citenamefont {Covey}, \citenamefont {Jin}
  \emph {et~al.}}]{zhu2014suppressing}%
  \BibitemOpen
  \bibfield  {author} {\bibinfo {author} {\bibfnamefont {B.}~\bibnamefont
  {Zhu}}, \bibinfo {author} {\bibfnamefont {B.}~\bibnamefont {Gadway}},
  \bibinfo {author} {\bibfnamefont {M.}~\bibnamefont {Foss-Feig}}, \bibinfo
  {author} {\bibfnamefont {J.}~\bibnamefont {Schachenmayer}}, \bibinfo {author}
  {\bibfnamefont {M.}~\bibnamefont {Wall}}, \bibinfo {author} {\bibfnamefont
  {K.~R.}\ \bibnamefont {Hazzard}}, \bibinfo {author} {\bibfnamefont
  {B.}~\bibnamefont {Yan}}, \bibinfo {author} {\bibfnamefont {S.~A.}\
  \bibnamefont {Moses}}, \bibinfo {author} {\bibfnamefont {J.~P.}\ \bibnamefont
  {Covey}}, \bibinfo {author} {\bibfnamefont {D.~S.}\ \bibnamefont {Jin}},
  \emph {et~al.},\ }\bibfield  {title} {\bibinfo {title} {Suppressing the loss
  of ultracold molecules via the continuous quantum zeno effect},\ }\href
  {https://link.aps.org/doi/10.1103/PhysRevLett.112.070404} {\bibfield
  {journal} {\bibinfo  {journal} {Physical Review Letters}\ }\textbf {\bibinfo
  {volume} {112}},\ \bibinfo {pages} {070404} (\bibinfo {year}
  {2014})}\BibitemShut {NoStop}%
\bibitem [{\citenamefont {Chaudhry}\ and\ \citenamefont
  {Gong}(2014)}]{chaudhry2014zeno}%
  \BibitemOpen
  \bibfield  {author} {\bibinfo {author} {\bibfnamefont {A.~Z.}\ \bibnamefont
  {Chaudhry}}\ and\ \bibinfo {author} {\bibfnamefont {J.}~\bibnamefont
  {Gong}},\ }\bibfield  {title} {\bibinfo {title} {Zeno and anti-zeno effects
  on dephasing},\ }\href {https://doi.org/10.1103/PhysRevA.90.012101}
  {\bibfield  {journal} {\bibinfo  {journal} {Phys. Rev. A}\ }\textbf {\bibinfo
  {volume} {90}},\ \bibinfo {pages} {012101} (\bibinfo {year}
  {2014})}\BibitemShut {NoStop}%
\bibitem [{\citenamefont {Facchi}\ and\ \citenamefont
  {Ligabo}(2010)}]{facchi2010quantum}%
  \BibitemOpen
  \bibfield  {author} {\bibinfo {author} {\bibfnamefont {P.}~\bibnamefont
  {Facchi}}\ and\ \bibinfo {author} {\bibfnamefont {M.}~\bibnamefont
  {Ligabo}},\ }\bibfield  {title} {\bibinfo {title} {Quantum zeno effect and
  dynamics},\ }\href {https://doi.org/10.1063/1.3290971} {\bibfield  {journal}
  {\bibinfo  {journal} {Journal of mathematical physics}\ }\textbf {\bibinfo
  {volume} {51}} (\bibinfo {year} {2010})}\BibitemShut {NoStop}%
\bibitem [{\citenamefont {Smerzi}(2012)}]{smerzi2012zeno}%
  \BibitemOpen
  \bibfield  {author} {\bibinfo {author} {\bibfnamefont {A.}~\bibnamefont
  {Smerzi}},\ }\bibfield  {title} {\bibinfo {title} {Zeno dynamics,
  indistinguishability of state, and entanglement},\ }\href
  {https://link.aps.org/doi/10.1103/PhysRevLett.109.150410} {\bibfield
  {journal} {\bibinfo  {journal} {Physical Review Letters}\ }\textbf {\bibinfo
  {volume} {109}},\ \bibinfo {pages} {150410} (\bibinfo {year}
  {2012})}\BibitemShut {NoStop}%
\bibitem [{\citenamefont {Facchi}\ \emph {et~al.}(2000)\citenamefont {Facchi},
  \citenamefont {Gorini}, \citenamefont {Marmo}, \citenamefont {Pascazio},\
  and\ \citenamefont {Sudarshan}}]{Facchi_2000}%
  \BibitemOpen
  \bibfield  {author} {\bibinfo {author} {\bibfnamefont {P.}~\bibnamefont
  {Facchi}}, \bibinfo {author} {\bibfnamefont {V.}~\bibnamefont {Gorini}},
  \bibinfo {author} {\bibfnamefont {G.}~\bibnamefont {Marmo}}, \bibinfo
  {author} {\bibfnamefont {S.}~\bibnamefont {Pascazio}},\ and\ \bibinfo
  {author} {\bibfnamefont {E.}~\bibnamefont {Sudarshan}},\ }\bibfield  {title}
  {\bibinfo {title} {Quantum zeno dynamics},\ }\href
  {https://doi.org/10.1016/s0375-9601(00)00566-1} {\bibfield  {journal}
  {\bibinfo  {journal} {Physics Letters A}\ }\textbf {\bibinfo {volume}
  {275}},\ \bibinfo {pages} {12–19} (\bibinfo {year} {2000})}\BibitemShut
  {NoStop}%
\bibitem [{\citenamefont {Xu}\ \emph {et~al.}(2011)\citenamefont {Xu},
  \citenamefont {Ai},\ and\ \citenamefont {Sun}}]{PhysRevA.83.022107}%
  \BibitemOpen
  \bibfield  {author} {\bibinfo {author} {\bibfnamefont {D.~Z.}\ \bibnamefont
  {Xu}}, \bibinfo {author} {\bibfnamefont {Q.}~\bibnamefont {Ai}},\ and\
  \bibinfo {author} {\bibfnamefont {C.~P.}\ \bibnamefont {Sun}},\ }\bibfield
  {title} {\bibinfo {title} {Dispersive-coupling-based quantum zeno effect in a
  cavity-qed system},\ }\href {https://doi.org/10.1103/PhysRevA.83.022107}
  {\bibfield  {journal} {\bibinfo  {journal} {Phys. Rev. A}\ }\textbf {\bibinfo
  {volume} {83}},\ \bibinfo {pages} {022107} (\bibinfo {year}
  {2011})}\BibitemShut {NoStop}%
\bibitem [{\citenamefont {Zhang}\ and\ \citenamefont {Xue}(2011)}]{Zhang_2011}%
  \BibitemOpen
  \bibfield  {author} {\bibinfo {author} {\bibfnamefont {Z.~T.}\ \bibnamefont
  {Zhang}}\ and\ \bibinfo {author} {\bibfnamefont {Z.~Y.}\ \bibnamefont
  {Xue}},\ }\bibfield  {title} {\bibinfo {title} {Demonstration of quantum zeno
  effect in a superconducting phase qubit},\ }\href
  {https://doi.org/10.1134/s0021364011060130} {\bibfield  {journal} {\bibinfo
  {journal} {JETP Letters}\ }\textbf {\bibinfo {volume} {93}},\ \bibinfo
  {pages} {349–353} (\bibinfo {year} {2011})}\BibitemShut {NoStop}%
\bibitem [{\citenamefont {McCusker}\ \emph {et~al.}(2013)\citenamefont
  {McCusker}, \citenamefont {Huang}, \citenamefont {Kowligy},\ and\
  \citenamefont {Kumar}}]{PhysRevLett.110.240403}%
  \BibitemOpen
  \bibfield  {author} {\bibinfo {author} {\bibfnamefont {K.~T.}\ \bibnamefont
  {McCusker}}, \bibinfo {author} {\bibfnamefont {Y.-P.}\ \bibnamefont {Huang}},
  \bibinfo {author} {\bibfnamefont {A.~S.}\ \bibnamefont {Kowligy}},\ and\
  \bibinfo {author} {\bibfnamefont {P.}~\bibnamefont {Kumar}},\ }\bibfield
  {title} {\bibinfo {title} {Experimental demonstration of interaction-free
  all-optical switching via the quantum zeno effect},\ }\href
  {https://doi.org/10.1103/PhysRevLett.110.240403} {\bibfield  {journal}
  {\bibinfo  {journal} {Phys. Rev. Lett.}\ }\textbf {\bibinfo {volume} {110}},\
  \bibinfo {pages} {240403} (\bibinfo {year} {2013})}\BibitemShut {NoStop}%
\bibitem [{\citenamefont {Kofman}\ and\ \citenamefont
  {Kurizki}(2000)}]{article}%
  \BibitemOpen
  \bibfield  {author} {\bibinfo {author} {\bibfnamefont {A.}~\bibnamefont
  {Kofman}}\ and\ \bibinfo {author} {\bibfnamefont {G.}~\bibnamefont
  {Kurizki}},\ }\bibfield  {title} {\bibinfo {title} {Acceleration of quantum
  decay processes by frequent observations},\ }\href
  {https://doi.org/10.1038/35014537} {\bibfield  {journal} {\bibinfo  {journal}
  {Nature}\ }\textbf {\bibinfo {volume} {405}},\ \bibinfo {pages} {546}
  (\bibinfo {year} {2000})}\BibitemShut {NoStop}%
\bibitem [{\citenamefont {Facchi}\ \emph {et~al.}(2001)\citenamefont {Facchi},
  \citenamefont {Nakazato},\ and\ \citenamefont
  {Pascazio}}]{PhysRevLett.86.2699}%
  \BibitemOpen
  \bibfield  {author} {\bibinfo {author} {\bibfnamefont {P.}~\bibnamefont
  {Facchi}}, \bibinfo {author} {\bibfnamefont {H.}~\bibnamefont {Nakazato}},\
  and\ \bibinfo {author} {\bibfnamefont {S.}~\bibnamefont {Pascazio}},\
  }\bibfield  {title} {\bibinfo {title} {From the quantum zeno to the inverse
  quantum zeno effect},\ }\href {https://doi.org/10.1103/PhysRevLett.86.2699}
  {\bibfield  {journal} {\bibinfo  {journal} {Phys. Rev. Lett.}\ }\textbf
  {\bibinfo {volume} {86}},\ \bibinfo {pages} {2699} (\bibinfo {year}
  {2001})}\BibitemShut {NoStop}%
\bibitem [{\citenamefont {Koshino}\ and\ \citenamefont
  {Shimizu}(2005)}]{Koshino_2005}%
  \BibitemOpen
  \bibfield  {author} {\bibinfo {author} {\bibfnamefont {K.}~\bibnamefont
  {Koshino}}\ and\ \bibinfo {author} {\bibfnamefont {A.}~\bibnamefont
  {Shimizu}},\ }\bibfield  {title} {\bibinfo {title} {Quantum zeno effect by
  general measurements},\ }\href
  {https://doi.org/10.1016/j.physrep.2005.03.001} {\bibfield  {journal}
  {\bibinfo  {journal} {Physics Reports}\ }\textbf {\bibinfo {volume} {412}},\
  \bibinfo {pages} {191–275} (\bibinfo {year} {2005})}\BibitemShut {NoStop}%
\bibitem [{\citenamefont {Slichter}\ \emph {et~al.}(2016)\citenamefont
  {Slichter}, \citenamefont {Müller}, \citenamefont {Vijay}, \citenamefont
  {Weber}, \citenamefont {Blais},\ and\ \citenamefont
  {Siddiqi}}]{Slichter_2016}%
  \BibitemOpen
  \bibfield  {author} {\bibinfo {author} {\bibfnamefont {D.~H.}\ \bibnamefont
  {Slichter}}, \bibinfo {author} {\bibfnamefont {C.}~\bibnamefont {Müller}},
  \bibinfo {author} {\bibfnamefont {R.}~\bibnamefont {Vijay}}, \bibinfo
  {author} {\bibfnamefont {S.~J.}\ \bibnamefont {Weber}}, \bibinfo {author}
  {\bibfnamefont {A.}~\bibnamefont {Blais}},\ and\ \bibinfo {author}
  {\bibfnamefont {I.}~\bibnamefont {Siddiqi}},\ }\bibfield  {title} {\bibinfo
  {title} {Quantum zeno effect in the strong measurement regime of circuit
  quantum electrodynamics},\ }\href
  {https://doi.org/10.1088/1367-2630/18/5/053031} {\bibfield  {journal}
  {\bibinfo  {journal} {New Journal of Physics}\ }\textbf {\bibinfo {volume}
  {18}},\ \bibinfo {pages} {053031} (\bibinfo {year} {2016})}\BibitemShut
  {NoStop}%
\bibitem [{\citenamefont {Cao}\ \emph {et~al.}(2012)\citenamefont {Cao},
  \citenamefont {Ai}, \citenamefont {Sun},\ and\ \citenamefont
  {Nori}}]{cao2012transition}%
  \BibitemOpen
  \bibfield  {author} {\bibinfo {author} {\bibfnamefont {X.}~\bibnamefont
  {Cao}}, \bibinfo {author} {\bibfnamefont {Q.}~\bibnamefont {Ai}}, \bibinfo
  {author} {\bibfnamefont {C.~P.}\ \bibnamefont {Sun}},\ and\ \bibinfo {author}
  {\bibfnamefont {F.}~\bibnamefont {Nori}},\ }\href@noop {} {\bibinfo {title}
  {The transition from quantum zeno to anti-zeno effects for a qubit in a
  cavity by modulating the cavity frequency}} (\bibinfo {year} {2012}),\
  \Eprint {https://arxiv.org/abs/1011.3862} {arXiv:1011.3862 [quant-ph]}
  \BibitemShut {NoStop}%
\bibitem [{\citenamefont {Chen}\ \emph {et~al.}(2010)\citenamefont {Chen},
  \citenamefont {Tsai},\ and\ \citenamefont {Bennett}}]{PhysRevB.81.115307}%
  \BibitemOpen
  \bibfield  {author} {\bibinfo {author} {\bibfnamefont {P.-W.}\ \bibnamefont
  {Chen}}, \bibinfo {author} {\bibfnamefont {D.-B.}\ \bibnamefont {Tsai}},\
  and\ \bibinfo {author} {\bibfnamefont {P.}~\bibnamefont {Bennett}},\
  }\bibfield  {title} {\bibinfo {title} {Quantum zeno and anti-zeno effect of a
  nanomechanical resonator measured by a point contact},\ }\href
  {https://doi.org/10.1103/PhysRevB.81.115307} {\bibfield  {journal} {\bibinfo
  {journal} {Phys. Rev. B}\ }\textbf {\bibinfo {volume} {81}},\ \bibinfo
  {pages} {115307} (\bibinfo {year} {2010})}\BibitemShut {NoStop}%
\bibitem [{\citenamefont {Barone}\ \emph {et~al.}(2004)\citenamefont {Barone},
  \citenamefont {Kurizki},\ and\ \citenamefont
  {Kofman}}]{PhysRevLett.92.200403}%
  \BibitemOpen
  \bibfield  {author} {\bibinfo {author} {\bibfnamefont {A.}~\bibnamefont
  {Barone}}, \bibinfo {author} {\bibfnamefont {G.}~\bibnamefont {Kurizki}},\
  and\ \bibinfo {author} {\bibfnamefont {A.~G.}\ \bibnamefont {Kofman}},\
  }\bibfield  {title} {\bibinfo {title} {Dynamical control of macroscopic
  quantum tunneling},\ }\href {https://doi.org/10.1103/PhysRevLett.92.200403}
  {\bibfield  {journal} {\bibinfo  {journal} {Phys. Rev. Lett.}\ }\textbf
  {\bibinfo {volume} {92}},\ \bibinfo {pages} {200403} (\bibinfo {year}
  {2004})}\BibitemShut {NoStop}%
\bibitem [{\citenamefont {Khalid}\ and\ \citenamefont
  {Chaudhry}(2019)}]{khalid2019quantum}%
  \BibitemOpen
  \bibfield  {author} {\bibinfo {author} {\bibfnamefont {B.}~\bibnamefont
  {Khalid}}\ and\ \bibinfo {author} {\bibfnamefont {A.~Z.}\ \bibnamefont
  {Chaudhry}},\ }\bibfield  {title} {\bibinfo {title} {The quantum zeno and
  anti-zeno effects: From weak to strong system-environment coupling},\ }\href
  {https://doi.org/10.1140/epjd/e2019-90681-3} {\bibfield  {journal} {\bibinfo
  {journal} {The European Physical Journal D}\ }\textbf {\bibinfo {volume}
  {73}},\ \bibinfo {pages} {1} (\bibinfo {year} {2019})}\BibitemShut {NoStop}%
\bibitem [{\citenamefont {Khan}\ \emph {et~al.}(2021)\citenamefont {Khan},
  \citenamefont {Soomro}, \citenamefont {Baig}, \citenamefont {Javed},\ and\
  \citenamefont {Chaudhry}}]{khan2021quantum}%
  \BibitemOpen
  \bibfield  {author} {\bibinfo {author} {\bibfnamefont {G.}~\bibnamefont
  {Khan}}, \bibinfo {author} {\bibfnamefont {H.}~\bibnamefont {Soomro}},
  \bibinfo {author} {\bibfnamefont {M.~U.}\ \bibnamefont {Baig}}, \bibinfo
  {author} {\bibfnamefont {I.}~\bibnamefont {Javed}},\ and\ \bibinfo {author}
  {\bibfnamefont {A.~Z.}\ \bibnamefont {Chaudhry}},\ }\bibfield  {title}
  {\bibinfo {title} {The quantum zeno and anti-zeno effects in the strong
  coupling regime},\ }\href {https://doi.org/10.48550/arXiv.2112.04850}
  {\bibfield  {journal} {\bibinfo  {journal} {arXiv preprint arXiv:2112.04850}\
  } (\bibinfo {year} {2021})}\BibitemShut {NoStop}%
\bibitem [{\citenamefont {Khan}\ \emph {et~al.}(2022)\citenamefont {Khan},
  \citenamefont {Soomro}, \citenamefont {Baig}, \citenamefont {Javed},\ and\
  \citenamefont {Chaudhry}}]{khan2022generalized}%
  \BibitemOpen
  \bibfield  {author} {\bibinfo {author} {\bibfnamefont {G.}~\bibnamefont
  {Khan}}, \bibinfo {author} {\bibfnamefont {H.}~\bibnamefont {Soomro}},
  \bibinfo {author} {\bibfnamefont {M.~U.}\ \bibnamefont {Baig}}, \bibinfo
  {author} {\bibfnamefont {I.}~\bibnamefont {Javed}},\ and\ \bibinfo {author}
  {\bibfnamefont {A.~Z.}\ \bibnamefont {Chaudhry}},\ }\bibfield  {title}
  {\bibinfo {title} {A generalized framework for the quantum zeno and anti-zeno
  effects in the strong coupling regime},\ }\href
  {https://doi.org/10.1038/s41598-022-23421-4} {\bibfield  {journal} {\bibinfo
  {journal} {Scientific Reports}\ }\textbf {\bibinfo {volume} {12}},\ \bibinfo
  {pages} {18652} (\bibinfo {year} {2022})}\BibitemShut {NoStop}%
\bibitem [{\citenamefont {Nielsen}\ and\ \citenamefont
  {Chuang}(2010)}]{nielsen2010quantum}%
  \BibitemOpen
  \bibfield  {author} {\bibinfo {author} {\bibfnamefont {M.~A.}\ \bibnamefont
  {Nielsen}}\ and\ \bibinfo {author} {\bibfnamefont {I.~L.}\ \bibnamefont
  {Chuang}},\ }\href@noop {} {\emph {\bibinfo {title} {Quantum computation and
  quantum information}}}\ (\bibinfo  {publisher} {Cambridge university press},\
  \bibinfo {year} {2010})\BibitemShut {NoStop}%
\bibitem [{\citenamefont {Beck}(2012)}]{beck2012quantum}%
  \BibitemOpen
  \bibfield  {author} {\bibinfo {author} {\bibfnamefont {M.}~\bibnamefont
  {Beck}},\ }\href@noop {} {\emph {\bibinfo {title} {Quantum mechanics: theory
  and experiment}}}\ (\bibinfo  {publisher} {Oxford University Press, USA},\
  \bibinfo {year} {2012})\BibitemShut {NoStop}%
\bibitem [{\citenamefont {Waseem}\ \emph {et~al.}(2020)\citenamefont {Waseem},
  \citenamefont {Anwar} \emph {et~al.}}]{waseem2020quantum}%
  \BibitemOpen
  \bibfield  {author} {\bibinfo {author} {\bibfnamefont {M.~H.}\ \bibnamefont
  {Waseem}}, \bibinfo {author} {\bibfnamefont {M.~S.}\ \bibnamefont {Anwar}},
  \emph {et~al.},\ }\href@noop {} {\emph {\bibinfo {title} {Quantum Mechanics
  in the Single Photon Laboratory}}}\ (\bibinfo  {publisher} {IOP Publishing},\
  \bibinfo {year} {2020})\BibitemShut {NoStop}%
\bibitem [{\citenamefont {Wootters}(1998)}]{wootters1998entanglement}%
  \BibitemOpen
  \bibfield  {author} {\bibinfo {author} {\bibfnamefont {W.~K.}\ \bibnamefont
  {Wootters}},\ }\bibfield  {title} {\bibinfo {title} {Entanglement of
  formation of an arbitrary state of two qubits},\ }\href
  {https://doi.org/10.1103/PhysRevLett.80.2245} {\bibfield  {journal} {\bibinfo
   {journal} {Phys. Rev. Lett.}\ }\textbf {\bibinfo {volume} {80}},\ \bibinfo
  {pages} {2245} (\bibinfo {year} {1998})}\BibitemShut {NoStop}%
\bibitem [{\citenamefont {Qi}\ \emph {et~al.}(2017)\citenamefont {Qi},
  \citenamefont {Gao},\ and\ \citenamefont {Yan}}]{qi2017measuring}%
  \BibitemOpen
  \bibfield  {author} {\bibinfo {author} {\bibfnamefont {X.}~\bibnamefont
  {Qi}}, \bibinfo {author} {\bibfnamefont {T.}~\bibnamefont {Gao}},\ and\
  \bibinfo {author} {\bibfnamefont {F.}~\bibnamefont {Yan}},\ }\bibfield
  {title} {\bibinfo {title} {Measuring coherence with entanglement
  concurrence},\ }\href {https://doi.org/10.1088/1751-8121/aa7638} {\bibfield
  {journal} {\bibinfo  {journal} {Journal of Physics A: Mathematical and
  Theoretical}\ }\textbf {\bibinfo {volume} {50}},\ \bibinfo {pages} {285301}
  (\bibinfo {year} {2017})}\BibitemShut {NoStop}%
\bibitem [{\citenamefont {Schlosshauer}(2007)}]{Schlosshauer:2014pgr}%
  \BibitemOpen
  \bibfield  {author} {\bibinfo {author} {\bibfnamefont {M.}~\bibnamefont
  {Schlosshauer}},\ }\href
  {https://doi.org/https://doi.org/10.1007/978-3-540-35775-9} {\emph {\bibinfo
  {title} {Decoherence and the Quantum-to-Classical Transition}}}\ (\bibinfo
  {publisher} {Springer Berlin, Heidelberg},\ \bibinfo {year}
  {2007})\BibitemShut {NoStop}%
\bibitem [{\citenamefont {Chaudhry}(2016)}]{chaudhry2016general}%
  \BibitemOpen
  \bibfield  {author} {\bibinfo {author} {\bibfnamefont {A.~Z.}\ \bibnamefont
  {Chaudhry}},\ }\bibfield  {title} {\bibinfo {title} {A general framework for
  the quantum zeno and anti-zeno effects},\ }\href
  {https://doi.org/10.1038/srep29497} {\bibfield  {journal} {\bibinfo
  {journal} {Scientific reports}\ }\textbf {\bibinfo {volume} {6}},\ \bibinfo
  {pages} {29497} (\bibinfo {year} {2016})}\BibitemShut {NoStop}%
\bibitem [{\citenamefont {Chaudhry}(2017)}]{chaudhry2017quantum}%
  \BibitemOpen
  \bibfield  {author} {\bibinfo {author} {\bibfnamefont {A.~Z.}\ \bibnamefont
  {Chaudhry}},\ }\bibfield  {title} {\bibinfo {title} {The quantum zeno and
  anti-zeno effects with strong system-environment coupling},\ }\href
  {https://rdcu.be/dtaW6} {\bibfield  {journal} {\bibinfo  {journal}
  {Scientific reports}\ }\textbf {\bibinfo {volume} {7}},\ \bibinfo {pages}
  {1741} (\bibinfo {year} {2017})}\BibitemShut {NoStop}%
\bibitem [{\citenamefont {Mirza}\ \emph {et~al.}(2023)\citenamefont {Mirza},
  \citenamefont {Jamil},\ and\ \citenamefont {Chaudhry}}]{mirza2023role}%
  \BibitemOpen
  \bibfield  {author} {\bibinfo {author} {\bibfnamefont {A.~R.}\ \bibnamefont
  {Mirza}}, \bibinfo {author} {\bibfnamefont {M.~N.}\ \bibnamefont {Jamil}},\
  and\ \bibinfo {author} {\bibfnamefont {A.~Z.}\ \bibnamefont {Chaudhry}},\
  }\bibfield  {title} {\bibinfo {title} {The role of initial system-environment
  correlations with a spin environment},\ }\href
  {https://doi.org/10.48550/arXiv.2301.07332} {\bibfield  {journal} {\bibinfo
  {journal} {arXiv preprint arXiv:2301.07332}\ } (\bibinfo {year}
  {2023})}\BibitemShut {NoStop}%
\bibitem [{\citenamefont {Breuer}\ and\ \citenamefont
  {Petruccione}(2002)}]{pet}%
  \BibitemOpen
  \bibfield  {author} {\bibinfo {author} {\bibfnamefont {H.-P.}\ \bibnamefont
  {Breuer}}\ and\ \bibinfo {author} {\bibfnamefont {F.}~\bibnamefont
  {Petruccione}},\ }\href@noop {} {\emph {\bibinfo {title} {The theory of open
  quantum systems}}}\ (\bibinfo  {publisher} {Oxford University Press, USA},\
  \bibinfo {year} {2002})\BibitemShut {NoStop}%
\bibitem [{\citenamefont {Magnus}(1954)}]{magnus}%
  \BibitemOpen
  \bibfield  {author} {\bibinfo {author} {\bibfnamefont {W.}~\bibnamefont
  {Magnus}},\ }\bibfield  {title} {\bibinfo {title} {On the exponential
  solution of differential equations for a linear operator},\ }\href
  {https://doi.org/10.1002/cpa.3160070404} {\bibfield  {journal} {\bibinfo
  {journal} {Communications on Pure and Applied Mathematics}\ }\textbf
  {\bibinfo {volume} {4}},\ \bibinfo {pages} {649} (\bibinfo {year}
  {1954})}\BibitemShut {NoStop}%
\bibitem [{\citenamefont {Mirza}\ \emph {et~al.}(2021)\citenamefont {Mirza},
  \citenamefont {Zia},\ and\ \citenamefont {Chaudhry}}]{PhysRevA.104.042205}%
  \BibitemOpen
  \bibfield  {author} {\bibinfo {author} {\bibfnamefont {A.~R.}\ \bibnamefont
  {Mirza}}, \bibinfo {author} {\bibfnamefont {M.}~\bibnamefont {Zia}},\ and\
  \bibinfo {author} {\bibfnamefont {A.~Z.}\ \bibnamefont {Chaudhry}},\
  }\bibfield  {title} {\bibinfo {title} {Master equation incorporating the
  system-environment correlations present in the joint equilibrium state},\
  }\href {https://doi.org/10.1103/PhysRevA.104.042205} {\bibfield  {journal}
  {\bibinfo  {journal} {Phys. Rev. A}\ }\textbf {\bibinfo {volume} {104}},\
  \bibinfo {pages} {042205} (\bibinfo {year} {2021})}\BibitemShut {NoStop}%
\bibitem [{\citenamefont {Blanes}\ \emph {et~al.}(2010)\citenamefont {Blanes},
  \citenamefont {Casas}, \citenamefont {Oteo},\ and\ \citenamefont
  {Ros}}]{Blanes_2010}%
  \BibitemOpen
  \bibfield  {author} {\bibinfo {author} {\bibfnamefont {S.}~\bibnamefont
  {Blanes}}, \bibinfo {author} {\bibfnamefont {F.}~\bibnamefont {Casas}},
  \bibinfo {author} {\bibfnamefont {J.~A.}\ \bibnamefont {Oteo}},\ and\
  \bibinfo {author} {\bibfnamefont {J.}~\bibnamefont {Ros}},\ }\bibfield
  {title} {\bibinfo {title} {A pedagogical approach to the magnus expansion},\
  }\href {https://doi.org/10.1088/0143-0807/31/4/020} {\bibfield  {journal}
  {\bibinfo  {journal} {European Journal of Physics}\ }\textbf {\bibinfo
  {volume} {31}},\ \bibinfo {pages} {907} (\bibinfo {year} {2010})}\BibitemShut
  {NoStop}%
\end{thebibliography}
\end{document}